%% file: orchestrating_HEP.tex
\documentclass[11pt,oneside]{article}
\usepackage[english]{babel}
\usepackage[utf8]{inputenc}
\usepackage[T1]{fontenc}
\usepackage{pmboxdraw}  
\usepackage[letterpaper,top=1in,bottom=1in,left=1in,right=1in,marginparwidth=1in]{geometry}

\usepackage{amsmath}
\usepackage{booktabs}
\usepackage{graphicx}
\graphicspath{{./figs/}}
\usepackage{url}
\usepackage[colorinlistoftodos]{todonotes}
\usepackage[colorlinks=true, allcolors=blue]{hyperref}
\usepackage{bbold}
\usepackage{listings} 
\usepackage{xcolor}
\usepackage{mdframed}
\usepackage{rotating}
\usepackage{lineno}

\newmdenv[
  backgroundcolor=gray!5,
  linecolor=black,
  linewidth=0.4pt,
  leftline=false,
  rightline=false,
  topline=true,
  bottomline=true]{graybox}
\newcommand{\reportnumber}{FERMILAB-PUB-25-0923-CSAID-ETD-T}
\usepackage[angle=0,scale=1,color=black,firstpage=true,opacity=1]{background}
\SetBgContents{\footnotesize\sf{\reportnumber}}
\SetBgPosition{current page.north east}
\SetBgHshift{-2in}
\SetBgVshift{-0.5in}
\usepackage{float}
\usepackage{mathtools}
\usepackage{braket}
\usepackage{slashed}
\usepackage[bottom]{footmisc}
\usepackage{pdfpages}
\usepackage{amsfonts,amsthm,amssymb}
\usepackage{type1cm}
\usepackage{lettrine}
\usepackage{sectsty}
\usepackage{titlesec}
\usepackage{xspace}
\usepackage{multirow}
\usepackage{cleveref}
\usepackage{tabularx}
\numberwithin{equation}{section}
\usepackage{setspace}
\usepackage[parfill]{parskip}
\usepackage{cite}

\input{orchestral}

\ifdefined\collatexdiffmode
  \IfFileExists{collatex.sty}{%
    \usepackage{collatex}%
    \collatexbaseline{archived/orchestrating_HEP_v1.tex}%
    \collatexskipenvs{verbatim,minted,lstlisting,Verbatim,orchestralsystemmessage,orchestralusermessage,orchestralagentmessage,orchestraltoolmessage,orchestraltoolmessageml,graybox}%
  }{}%
\fi

\newcommand{\pkgname}{HEPTAPOD\xspace}

\title{\textbf{\pkgname: Orchestrating High Energy Physics Workflows Towards Autonomous Agency}}

\author{Tony Menzo$^{1,2}$\footnote{Co-first author, corresponding author: \texttt{amenzo@ua.edu}},\, Alexander Roman\footnote{Co-first author.},\,  Sergei Gleyzer, Konstantin Matchev \\
{$^1$}\textit{Department of Physics and Astronomy, University of Alabama, Tuscaloosa, AL 35487, USA}\\[12pt]
George T. Fleming, Stefan H\"{o}che, Stephen Mrenna, Prasanth Shyamsundar\\
{$^2$}\textit{Fermi National Accelerator Laboratory, Batavia, IL 60510, USA}
}

\date{July 17, 2026}

\begin{document}
\maketitle

\begin{center}
\begin{minipage}{13cm}

\section*{\centering Abstract}
Many workflows in high-energy-physics (HEP) stand to benefit from recent advances in transformer-based large language models (LLMs). 
While early applications of LLMs focused on text generation and code completion, modern LLMs now support \emph{orchestrated agency}: the coordinated execution of complex, multi-step tasks through tool use, structured context, and iterative reasoning. 
We introduce the \textbf{HEP} \textbf{T}oolkit for \textbf{A}gentic \textbf{P}lanning, \textbf{O}rchestration, and \textbf{D}eployment (\pkgname), an orchestration framework designed to integrate external LLMs into general HEP workflows spanning theoretical calculations, simulation, and data analysis. 
The framework enables LLMs to interface with domain-specific tools and to construct and manage diverse HEP pipelines while preserving transparency, reproducibility, and human oversight.
To demonstrate these capabilities, we present a representative case study in the context of a Beyond the Standard Model (BSM) Monte Carlo signal validation that spans model generation, event simulation, and analysis within an established, reproducible workflow.  
\pkgname provides a structured and auditable layer between human researchers, LLMs, and computational infrastructure, establishing a foundation for transparent, human-in-the-loop systems.

\end{minipage}
\end{center}
\vspace*{1cm}

\newpage
\tableofcontents
\newpage
\section{Introduction}
\label{sec:introduction}

Large language models (LLMs) have evolved from text generators into general-purpose computational interfaces, capable of planning and executing multi-step workflows through structured prompting and function calls \cite{openai2024gpt4technicalreport,bubeck2023sparks,zhao2023survey,yin2023multimodal,minaee2025largelanguagemodelssurvey,grattafiori2024llama3herdmodels}. 
Their capabilities appear to scale predictably with model size, dataset quality, and training compute, following empirical scaling laws that remain robust across architectures and domains \cite{kaplan2020scaling,hoffmann2022training}. 
This progression has introduced the notion of \emph{LLM agency}, in which models operate autonomously or cooperatively to complete complex objectives \cite{guo2024largelanguagemodelbased,tran2025multiagentcollaborationmechanismssurvey,li2024survey_on_llm_based_multi_agent_systems}. 
As investments in large-scale compute infrastructure continue, these systems are expected to improve in reasoning, tool use, and generalization, suggesting a near-term transition from passive language processors to reliable orchestrators of scientific workflows \cite{wei2025aiscienceagenticscience,jiang2025agenticscimlcollaborativemultiagentsystems}. This momentum is already visible across the sciences, with agentic systems applied to software engineering \cite{metagpt,sweagent}, chemistry \cite{chemcrow,coscientist}, economics \cite{econagent}, and cosmology \cite{Laverick:2024mas,Xu:2025cmbagent,Moss:2025aicosmologist}.

High-energy physics (HEP) presents a natural testing ground for these capabilities. 
Modern theory and phenomenology rely on sophisticated computational stacks spanning symbolic algebra, matrix-element generation, parton showering and hadronization, detector simulation, and numerical analysis, all of which must operate in coordinated, reproducible sequences \cite{Ask:2012sm}. 
Yet most pipelines remain manually managed, requiring researchers to orchestrate long chains of interdependent calculations, propagate parameters across tools, interpret heterogeneous outputs, and debug failures at any stage.

Concurrently, machine learning (ML) methods have reshaped HEP research across simulation \cite{Butter:2022rso} (including phase-space integration and event unweighting \cite{Bothmann:2020ywa,Gao:2020zvv}, hadronization modeling \cite{Ilten:2022jfm,Bierlich:2023zzd,Bierlich:2024xzg,Assi:2025avy,Butter:2025wxn}, and fast detector simulation using generative models \cite{CaloChallenge2024}), theory inputs such as parton distribution function determinations
\cite{NNPDF:2021njg,NNPDF:2021uiq}, event reconstruction \cite{ExaTrkX:2020nyf}, triggering \cite{Iiyama:2020wap,Duarte:2019fta}, anomaly detection \cite{Aarrestad:2021oeb,Kasieczka:2021xcg}, unfolding \cite{Andreassen:2019cjw,Bellagente:2020piv}, and analysis \cite{Albertsson:2018maf,Guest:2018yhq}. 
Transformers in particular have become standard architectures for particle-cloud reasoning in jet physics, with applications ranging from task-specific jet classification \cite{Qu:2019gqs,Qu:2022mxj} to autoregressive generative modeling of jet radiation \cite{Butter:2023fov} and more recently to pretrained foundation models for jet representations \cite{Birk:2024knn,Golling:2024abg}.
Related but more general efforts have begun to explore pretrained transformer foundation models for collider-level data \cite{Bardhan:2025icr}.
Early applications of transformer-based LLMs in HEP focused on code generation, documentation, and integration with existing software infrastructures \cite{Zhang:2024kws,Atif:2025mkh}. Within HEP itself, agentic frameworks have likewise begun to appear: \texttt{ArgoLOOM}, a single-agent orchestrator with a retrieval-augmented knowledge base of the literature that iteratively invokes steering scripts wrapping external physics codes such as \textsc{CLASS} and \textsc{MadGraph} across cosmology, collider, and nuclear physics \cite{Bakshi:2025fgx}; \textit{Agents of Discovery}, a team of collaborating LLM agents that performs anomaly-detection analyses on LHC Olympics data \cite{Diefenbacher:2025aod}; and \texttt{LLM4HEP}, which pairs an LLM supervisor--coder agent with the \textsc{Snakemake} workflow manager to autonomously generate, execute, and correct the analysis code for a Higgs-boson cross-section measurement on ATLAS Open Data \cite{GendreauDistler:2025hep}.

We introduce the \textbf{HEP} \textbf{T}oolkit for \textbf{A}gentic \textbf{P}lanning, \textbf{O}rchestration, and \textbf{D}eployment (\pkgname)\footnote{Not to be confused with the image autoregressive model for language modeling on visual signals introduced in \cite{zhu2025heptapodlanguagemodelingvisual}. That model, also referred to as HEPTAPOD, is named after the alien species in the film Arrival (2016).}, an orchestration framework that integrates LLMs directly into HEP theory pipelines. The agentic execution layer is implemented using the {\sc Orchestral AI} framework~\cite{orchestral-ai}.\footnote{A companion paper showcases an analogous application of {\sc Orchestral AI} for the analysis and simulation of exoplanet transit spectroscopy \cite{panek2026asteragenticscience}.}
{\sc Orchestral} is a general-purpose LLM-provider-agnostic agent-orchestration harness that manages the evolving conversation context, exposes tool schemas to the model, and drives the reasoning--execution loop, while delegating language understanding to an external LLM accessed either through a hosted API (e.g.\ OpenAI or Anthropic) or a locally served open-weights model (e.g.\ via Ollama). In this respect it is architecturally analogous to general-purpose coding harnesses such as Claude Code, OpenAI Codex, or opencode, but \pkgname configures it around domain-specific scientific tools and schema-validated physics artifacts rather than software-engineering tasks. Like those frameworks, {\sc Orchestral AI} and \pkgname ship no language model of their own; an external LLM must be supplied by the user.
Rather than pursuing full autonomy, \pkgname emphasizes human-in-the-loop control: the LLM acts as a workflow coordinator that invokes domain-specific tools, interprets intermediate outputs, and proposes next steps, while researchers retain oversight and validation.
This provides the necessary structure so that increasingly-capable scientific agents can operate safely and reproducibly within established computational workflows.

The HEPTAPOD framework is well aligned with emerging initiatives that seek to leverage agentic AI systems for discovery-driven science.\footnote{In particular, it supports the objectives of the recently launched Genesis Mission, including the Discovery Science challenge of ``Understanding the universe, from quarks to cosmos'' \cite{DiscoveryScience}.} 
Within this landscape, \pkgname differs from existing agentic HEP frameworks in its reliance on \emph{schema-validated tools} as the basis for agentic orchestration of HEP tasks, with recurring operations packaged as typed interfaces whose inputs and outputs are validated before and after execution. 
These provide the agent with verifiable, composable building blocks from which to assemble multi-stage workflows, complemented by direct code generation where a fixed interface is unwarranted. The contribution of this paper and accompanying codebase is the orchestration infrastructure itself, implemented, documented, and exercised in full by the worked example and quantitative benchmark that follow. By abstracting theory tasks into modular, schema-validated tools, \pkgname enables applications ranging from effective-field-theory construction to automated event-generation pipelines and symbolic-numerical translation.
More broadly, it provides a blueprint for how agentic LLMs can augment scientific reasoning without displacing human judgment, establishing a foundation for reproducible, adaptive, and interpretable AI-assisted research in theoretical physics.

The rest of this paper is organized as follows.
\Cref{sec:example} introduces a representative leptoquark workflow that will serve as a running example 
throughout the paper. 
\Cref{sec:architecture} describes the orchestration architecture underlying \pkgname, including schema-validated tools, LLM-friendly event formats, run-card-driven coordination, and the distinction between orchestration and traditional scripting approaches. 
\Cref{sec:example_full} demonstrates the complete framework through a BSM leptoquark parameter scan and signal validation.
\Cref{sec:benchmarking} benchmarks the reliability of the framework on this workflow, comparing agent performance with and without the domain-specific tools.
\Cref{sec:conclusions} concludes with a discussion of future work and perspectives on increasingly agentic scientific systems.

\section{An example workflow}
\label{sec:example}

To motivate the design goals of \pkgname and to illustrate the challenges involved in coordinating multi-stage high-energy-physics workflows, we introduce a representative example that will serve as a running case study. 
Specifically, we consider the Monte Carlo simulation pipeline required to study new BSM models that are not available by default in general-purpose event generators. 
In this situation, model implementation, event generation, and downstream simulation must be manually composed across multiple software packages. 
The design and automation of this pipeline was the subject of a series of international workshops, Monte Carlo Tools for Beyond the Standard Model Physics (MC4BSM), which ran in the 2000's and 2010's \cite{MC4BSM}. The example considered here is very close to the test case study demonstrated in the MC4BSM tutorials \cite{Ask:2012sm}.

For concreteness, we consider a scalar leptoquark~\cite{Dorsner:2016wpm}
\begin{equation}
S_1 \sim (\bar{\mathbf{3}}, \mathbf{1}, 1/3),
\end{equation}
with a simplified interaction Lagrangian
\begin{equation}
\mathcal{L} \supset 
(D_\mu S_1)^\dagger (D^\mu S_1)
+ y_{ij}\,\overline{u^c_{Ri}}\, e_{Rj}\, S_1
+ \mathrm{h.c.}
\end{equation}
Assuming diagonal Yukawa couplings, pair production proceeds dominantly through QCD,
\begin{equation}
pp \to S_1 S_1^\dagger \to (\ell^+ j)(\ell^- j),
\end{equation}
yielding a characteristic \(2\ell + 2j\) topology. 
The task is to perform MC signal generation over a scan of the leptoquark mass $m_{S_1}$ and predict the distribution of the reconstructed leptoquark mass $m^{\text{min}}_{\text{LQ}}$, which for concreteness is taken as the smaller of the two reconstructed masses of the leptoquark candidates $m_{\text{LQ}}^{(1)}$ and $m_{\text{LQ}}^{(2)}$
\begin{equation}\label{eq:mLQmin}
m^{\min}_{\text{LQ}} \;=\;
\min\left\{m_{\text{LQ}}^{(1)}, m_{\text{LQ}}^{(2)}\right\},
\end{equation}
where the lepton-jet assignments are selected to minimize the absolute difference
$\left|m_{\text{LQ}}^{(1)} - \, m_{\text{LQ}}^{(2)}\right|$.
A typical phenomenological study varies $m_{S_1}$ over a set of benchmark values, e.g., ($1.0, 1.5, 2.0~\mathrm{TeV}$), generating and analyzing events for each mass point. 
This requires coordinated propagation of model parameters, consistent run-card configuration across tools, and synchronized analysis logic across the scan. 
The scan is simple, yet has the essential structure of a multi-stage HEP workflow: configuration must propagate consistently across several tools, intermediate files must be handled carefully, and analysis logic must remain synchronized across scan points. 

Although not included here, realistic validation pipelines must also include the dominant SM backgrounds to this signature. For the $(\ell j)(\ell j)$ final state, the most relevant backgrounds are
\begin{itemize}
    \item Drell--Yan + jets: $pp \to Z/\gamma^\ast + \mathrm{jets} \to \ell^+\ell^- + \mathrm{jets}$,
    \item Top-quark pair production: $pp \to t\bar{t} \to (b\ell^+\nu)(\bar{b}\ell^-\bar{\nu})$,
    \item Diboson production: $WW$, $WZ$, or $ZZ$ events with associated jets.
\end{itemize}
Each background requires its own run card, generation chain, and analysis flow, producing multiple parallel branches that must remain consistent. 
These branches are straightforward to implement within the orchestration framework because they reuse the same tool chain with different run-card inputs.
Here we focus solely on the signal to illustrate the orchestration framework.

\begin{figure}[t!]
    \centering
    \includegraphics[width=0.97\textwidth]{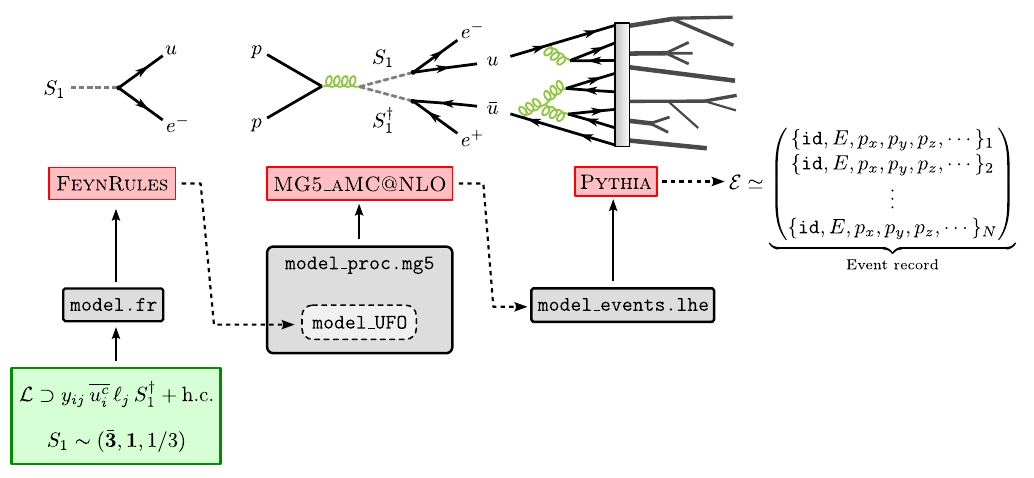}
    \caption{
    One example of a multi-phase software workflow for a BSM leptoquark study. 
    A \texttt{model.fr} file is implemented in FeynRules and exported as a UFO directory (\texttt{model\_UFO/}). The UFO is passed to \textsc{MG5\_aMC@NLO}, which loads the process card \texttt{model\_proc.mg5} and generates parton-level events written to \texttt{model\_events.lhe}. 
    These events are then passed to \textsc{Pythia} for showering and hadronization, producing an event record $\mathcal{E}$ containing particle identifiers and four-momenta $\{\,p_x, p_y, p_z, E, \texttt{id}, \ldots\}$. 
    }\label{fig:hep_workflow}
\end{figure}

For this example, a complete HEP workflow can be instantiated in several ways. One representative realization is considered here (see \cref{fig:hep_workflow}), though comparable pipelines using, for example, \textsc{Sherpa} \cite{Sherpa:2024mfk} follow the same basic structure:
\begin{enumerate}
    \item \textbf{Model generation:} convert the \textsc{FeynRules} \cite{Alloul:2013bka} model file into a Universal FeynRules Output (UFO) directory.
    \item \textbf{Parton-level event generation:} use \textsc{MadGraph5\_aMC@NLO} \cite{Alwall:2014hca} to generate events for each mass point specified in the run card via a user-defined parameter scan.
    \item \textbf{Showering and hadronization:} use \textsc{Pythia} 
    \cite{Bierlich:2022pfr} to process the LHE events into particle-level final states.
    \item \textbf{Jet reconstruction:} apply anti-\(k_T\) or Cambridge/Aachen clustering via \textsc{FastJet} \cite{Cacciari:2011ma}.
    \item \textbf{Resonance reconstruction:} pair leptons and jets, compute $m_{\text{LQ}}^{(1)}, m_{\text{LQ}}^{(2)}$, and derive $m^{\min}_{\text{LQ}}$.
\end{enumerate}

Careful management of run cards, file paths, tool configurations, and intermediate data formats is necessary to execute this sequence reliably. 
Manual pipelines are time-intensive and error-prone. 
In later sections we return to this example and show how \pkgname treats each phase as a schema-validated tool, propagates scan metadata through the workflow automatically, and uses run cards as explicit coordination boundaries. 
This allows an LLM to orchestrate the full analysis while preserving transparency, reproducibility, and human oversight.

\section{Orchestration architecture and philosophy}
\label{sec:architecture}

\begin{figure}[t!]
    \centering
    \includegraphics[width=0.81\textwidth]{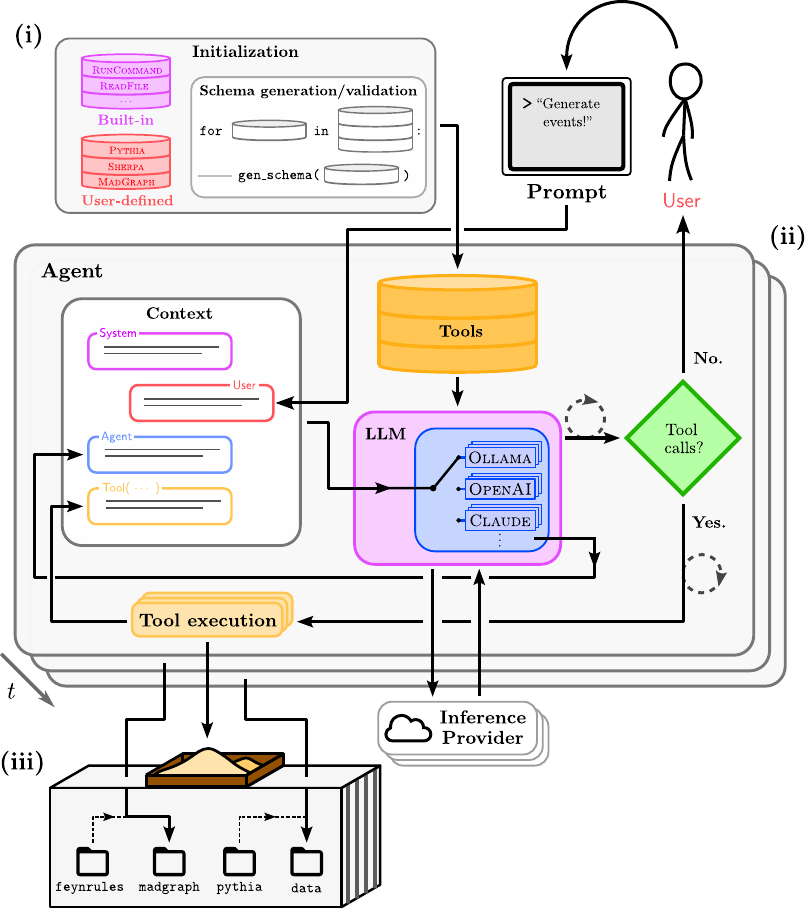}
    \caption{
    High-level architecture of \pkgname, a HEP-focused agent-orchestration framework built using the \textsc{Orchestral AI} orchestration engine \cite{orchestral-ai}. The system consists of three interacting components: \textbf{(i)} a database of schema-validated tools that expose domain-specific HEP capabilities via structured inputs and outputs; \textbf{(ii)} an agent-orchestration layer that maintains the evolving context and performs LLM-based reasoning, optionally routing inference across multiple providers; and \textbf{(iii)} a sandboxed tool-execution engine that interfaces with external HEP software and data utilities.
    Each agent interaction proceeds within an initialized sandboxed workspace, with the agent iteratively reasoning over the evolving context.
    At each step, the agent determines whether a tool invocation is required; when invoked, tools are executed outside the LLM within the sandbox, and their structured outputs (e.g.\ JSON artifacts) are injected back into the context before the next reasoning step. This closed-loop reasoning–execution process enables reliable, multi-step HEP workflows while enforcing a strict separation between language-model reasoning and external software execution.
    }\label{fig:heptapod_workflow}
\end{figure}

The central objective of \pkgname is to facilitate and enable large language models (LLMs) to reliably coordinate high-energy-physics (HEP) workflows while preserving transparency, reproducibility, and human oversight. 
The leptoquark workflow introduced in \cref{sec:example} highlights the central challenge in automating HEP analyses: while each individual phase of the pipeline is well supported by mature software packages, the coordination of these components remains largely manual. 
Researchers must construct run cards, manage intermediate files, ensure parameter consistency, and sequence tool execution correctly across multiple stages. 
These tasks are labor-intensive and error-prone, especially when extended across parameter scans or repeated for large model spaces.
The following subsections outline the architectural features of \pkgname that address these coordination challenges. 
A high-level view of the HEPTAPOD architecture and its interaction between agents, tools, and external HEP software is shown in \cref{fig:heptapod_workflow}.

\subsection{Schema-validated tools}
\label{subsec:schematools}

Each scientific capability in \pkgname is exposed to the agent as a \emph{tool}: a Python class with a structured JSON schema defining its inputs and outputs.
Tools serve as the fundamental interface between the agent and external HEP software. 
Within the orchestration framework, tools play three roles: 1.~to formalize how the agent invokes domain-specific capabilities, 2.~to return structured outputs that downstream steps can reliably reuse, and 3.~to define clear execution boundaries that support reproducibility and human oversight. 

Tool schemas distinguish between \emph{runtime fields}, which the agent provides when invoking the tool, and \emph{state fields}, which are injected by the orchestration layer and not visible to the agent. 
Runtime fields correspond to the scientific and programmatic degrees of freedom, like relative paths, algorithmic configurations, and parameter choices. 
State fields supply environmental information such as workspace locations, internal defaults (\textit{e.g.}~dependency paths), or metadata from previous phases.
This separation keeps the agent's reasoning at the physics level while insulating it from unnecessary overhead from the details of the filesystem or workspace (\textit{a.k.a.}~sandbox).

A tool's schema is defined directly by its class annotations, from which the orchestration layer derives validation rules that enforce argument conventions and ensure that all values exchanged between the agent and the execution environment are checked prior to tool execution. 
For example, a simplified \textsc{Pythia} interface might declare:
\begin{graybox}
    \begin{verbatim}
class PythiaFromRunCardTool(BaseTool):
    """
    Generate hadron-level events using Pythia8 from a .cmnd card.

    Inputs (runtime):
      - command_card: path to Pythia8 .cmnd configuration file
      - nevents: number of events to generate
      - seed: optional random seed
      ...

    Behavior:
      1. Initialize Pythia8 with the provided run card
      2. Generate nevents events with optional fixed seed
      3. Write structured event records to JSONL format
      ...

    Output (JSON):
      {"status": "ok", "events_jsonl": "<path>", ...}
    """
    # Runtime fields
    command_card: str = RuntimeField(
        description="Path to Pythia8 .cmnd configuration file")
    nevents: int = RuntimeField(
        description="Number of events to generate")
    seed: Optional[int] = RuntimeField(
        default=None, description="Random seed (optional)")
    \end{verbatim}
\end{graybox}

An agent invoking this tool submits a structured tool call, whose runtime arguments are validated against their declared types before the tool body runs. The \texttt{BaseTool} layer, a \texttt{Pydantic} model, re-instantiates the tool from its schema and rejects any type-inconsistent call as a structured error. This check runs inside \pkgname's own process, independently of which inference backend produced the call. Some backends can additionally constrain generation to a grammar compiled from the tool schema, so that malformed calls are never emitted in the first place. To remain provider-agnostic, however, \pkgname does not rely on such constrained decoding.
When failures do arise from external software (e.g. \textsc{FeynRules}, \textsc{MadGraph}, \textsc{Pythia}), tools return structured error messages that the agent can interpret and act upon.
This provides predictable failure modes in place of unstructured logs.
The translation from raw process output to a structured error message is performed deterministically by the tool wrapper (the \texttt{BaseTool} layer, Appendix~\ref{app:tools}) rather than by the LLM. When a tool fails, the wrapper detects the failure (a nonzero exit status from an external code, an unmet precondition, or a caught exception) and returns a consistently formatted error message through \texttt{BaseTool}'s \texttt{format\_error} routine, naming the failure together with any context useful for diagnosis, such as the offending field or a pointer to a log for verbose external output. This message is appended to the active context and sent back to the model at the next reasoning step, so the agent \emph{interprets} a compact, structured error rather than \emph{producing} one from free-form logs. Because \texttt{format\_error} is inherited from \texttt{BaseTool}, every user-defined tool reports errors in the same form automatically.

As a design convention, tools also return structured JSON outputs. This ensures that each tool produces data in a form that downstream steps can reuse without ad hoc parsing or heuristics.
The output schema specifies the fields and their types, which guarantees that intermediate products such as UFO directories, LHE file paths, jet collections, or analysis summaries have a predictable structure. 
This removes ambiguity at tool boundaries and provides the agent with machine-checkable data structures rather than free-form text, which is essential for coordinating multi-step scientific workflows.
An example tool invocation and the associated JSON output for the \texttt{PythiaFromRunCardTool} is shown below:
\vspace{-0.2in}
\begin{minted}[escapeinside=||]{text}
PythiaFromRunCardTool(
    command_card="cards/run_ppbar.cmnd",
    nevents=5000,
    seed=12345
)
|$\to$| {"status": "ok", "events_jsonl": "data/run_01/events.jsonl", ...}
\end{minted}

Crucially, in addition to the schema itself, the \emph{tool docstring} (the Python docstring, a free-form text description attached to a function or class that explains what it does and how to use it) plays a critical role in guiding the agent's behavior. 
Each tool class carries a domain-specific docstring that describes its scientific purpose, the meaning of its runtime fields, expected behavior, and the semantics of its outputs.
This docstring functions as the tool's prompt: when the agent considers invoking the tool, the orchestration layer provides both the schema and the docstring, giving the agent a structured interface (the JSON schema) and a semantic interface (the natural-language description).
The tool docstring is extracted and included in the tool specification sent with every LLM API call, ensuring that the agent has immediate access to semantic context needed to understand what parameters to provide and how to interpret the returned fields without relying on conversational memory.
The combination is essential.
The schema tells the agent what inputs are allowed; the docstring tells the agent why and when the tool should be used.

Because outputs are returned as structured JSON, the tool docstring also defines the interpretation of those fields.
For example, the \textsc{MadGraph} tool docstring specifies that when a parameter scan is detected, the returned \texttt{runs} array contains objects with \texttt{run\_id}, \texttt{lhe\_file}, \texttt{scan\_params}, and \texttt{cross\_section} fields, where all file paths are relative to the sandbox.
This coupling between semantic guidance and machine-checkable structure enables the agent to reason about workflows without relying on free-form text generation or fragile pattern matching.
The agent uses the docstring to decide \emph{which} tool to invoke and the schema to decide \emph{how} to invoke it.

Together, the schema and the docstring act synergistically: the docstring communicates intent and scientific context; the schema enforces shape, type, and validity.
This is ultimately what allows tools to operate as reliable, composable units within multi-step workflows.

\subsection{Tools versus free-form generation}
\label{subsec:goodtool}

A natural question is when a capability should be exposed as a dedicated tool versus left to free-form, in-line code generation by the agent. \pkgname does not treat every operation as requiring a tool; rather, it reserves tools for operations that satisfy one or more of the following criteria:

\paragraph{Recurrence.} The operation appears across many analyses rather than only the workflow at hand. Selecting the hardest $N$ objects, computing $p_T$ or $\Delta R$, filtering by PDG ID, and reconstructing invariant masses are ubiquitous across BSM searches, SM measurements, and detector studies, and so are worth packaging once and reusing.

\paragraph{Schema-stable input/output.} The operation uses or produces artifacts whose schema is stable enough to support reliable downstream use (e.g.\ event records, jet collections, scan metadata). A stable contract is what allows tools to compose without ad hoc glue code.

\paragraph{Auditability at a glance.} Correctness should be verifiable from the call signature alone, without reading the body of generated code on every invocation. This is central to the human-in-the-loop, reviewable design of \pkgname.

\paragraph{Wraps external software.} Operations that interface with external HEP codes (\textsc{MadGraph}, \textsc{Pythia}, \textsc{FastJet}, \textsc{FeynRules}) benefit from absorbing configuration complexity and failure handling behind a stable, validated contract.

\paragraph{Determinism and reproducibility.} Operations whose results must be identical across runs and across studies are better served by a fixed, versioned implementation than by regenerated code.

Conversely, in-line LLM execution or free-form code generation is the appropriate mode for one-off operations local to a single workflow and for exploratory or custom analyses whose required output cannot be anticipated, where flexibility outweighs the auditability and recurrence advantages of a fixed interface. For example, a plot whose form is not known in advance has nothing stable to wrap and is best left to free-form generation, whereas a common, standardized plot is better implemented as a tool (or embedded into the tool that generates the underlying data itself). Such free-form generation carries a verification debt, since, unlike a tool that is validated once and trusted thereafter, generated code must be re-checked on each run. The leptoquark workflow in \cref{sec:example_full} exercises both modes, with the analysis steps invoking schema-validated tools while the final $m^{\min}_{\text{LQ}}$ histogram is produced by agent-generated Python (\cref{subsec:example_convo}) rather than a dedicated plotting tool.

\subsection{LLM-compatible event data structures}
\label{subsec:evtjsonl}

Effective orchestration requires that intermediate physics data, particularly event-level outputs, be stored in a form that is both machine-actionable and directly interpretable by a language model, enabling the agent to participate meaningfully in validation, debugging, and exploratory analysis.
For this reason, \pkgname introduces a structured, text-based representation for intermediate event data, based on a line-delimited JSON (JSONL) format referred to throughout as the \texttt{evtjsonl} event format.

Traditional HEP event containers and event-record formats such as \textsc{Root} \cite{Antcheva:2009zz}, \textsc{HDF5}, \textsc{HepMC} \cite{Buckley:2019xhk} and \textsc{LHE} \cite{Alwall:2006yp} are optimized for storage density and I/O throughput in large scale simulation and analysis workflows.
While well suited to large-scale production and offline analysis, these formats are opaque to language models: they require specialized libraries for deserialization and expose internal structure only indirectly.
As a result, an LLM cannot inspect individual events, validate intermediate outputs, or reason about event content without invoking external tools.
In contrast, the JSONL representation encodes each event as a self-contained record that can be examined directly within the conversational context.
This allows the agent to inspect representative events, verify kinematic properties, and respond to qualitative queries about the data without leaving the orchestration layer. This legibility also supports a semantic tier of validation that complements the structural schema checks of \cref{subsec:schematools}, with the agent flagging content-level anomalies, such as implausible multiplicities or kinematically inconsistent four-vectors, that structural validation alone cannot detect.

The \texttt{evtjsonl} format standardizes this representation across all workflow stages while remaining intentionally extensible for future use cases (\textit{e.g.}~long-lived particle production, \textit{etc.}).
Each event record contains an \texttt{event\_id} field and a \texttt{data} object, which holds either a \texttt{particles} array for particle-level events or a \texttt{jets} array for reconstructed jet collections.
Particle objects include kinematic four-vectors $(p_x, p_y, p_z, E)$, mass $m$, PDG identifier \texttt{id}, and status codes; jet objects carry four-vectors and jet-specific metadata.
This uniform structure allows tools to read event data predictably: all event generators, shower algorithms, jet clustering tools, and analysis selections produce and expect the same format, eliminating the need for ad hoc parsing or format conversions between the orchestration layer and downstream tools. We emphasize that this does not eliminate parsing of the external codes' native formats but \emph{localizes} it inside each tool wrapper, behind the uniform \texttt{evtjsonl} interface, instead of repeating it at every stage boundary.

The line-delimited structure of JSONL also enables incremental processing.
Because each line is an independent record, tools can iterate through events sequentially, processing and writing results one event at a time without loading entire datasets into memory.
The format also simplifies parallelization and filtering: events can be distributed across workers or selected on-the-fly using standard command-line tools.
This design choice reflects a broader principle within \pkgname: intermediate data formats should be interpretable not only by the orchestration layer and downstream analysis tools, but also by the agent itself.

While JSONL sacrifices storage compactness relative to compressed formats, \pkgname mitigates this through interoperability tools.
\texttt{LHEToJSONLTool} imports Les Houches events from matrix-element generators; \texttt{EventJSONLToNumpyTool} and \texttt{JetsJSONLToNumpyTool} export to zero-padded NumPy arrays compatible with vectorized analysis libraries and machine-learning frameworks such as JAX, PyTorch, and TensorFlow.
The architecture also supports emitting a standard \textsc{HepMC} \cite{Buckley:2019xhk} file in parallel at the shower/hadronization boundary, so that a \texttt{.hepmc} record is retained as a conventional staging checkpoint while \texttt{evtjsonl} serves the orchestration layer. A complementary tool that opens such a file and returns summary content to the agent, using the libraries already present in the pipeline, is a natural planned addition (see \cref{sec:conclusions}).
This hybrid approach retains the agent-facing interpretability of JSONL during orchestration while enabling seamless integration with the broader HEP and ML ecosystems at analysis endpoints.

\subsection{Run-card-driven orchestration}
\label{subsec:runcards}

Most major HEP collider simulation frameworks, including \textsc{FeynRules} \cite{Alloul:2013bka}, \textsc{MadGraph} \cite{Alwall:2011uj}, \textsc{Pythia} \cite{Bierlich:2022pfr}, \textsc{Sherpa} \cite{Sherpa:2024mfk}, \textsc{Herwig} \cite{Bewick:2023tfi}, \textsc{Delphes} \cite{deFavereau:2013fsa}, \textsc{Geant4} \cite{Allison:2016lfl}, and \textsc{Rivet} \cite{Bierlich:2019rhm}, are configured via human-readable run cards or model files. 
These declarative configuration files specify model parameters, beam conditions, detector settings, event-selection criteria, and analysis instructions. 
\pkgname adopts run cards as the canonical interface between the agent and external HEP codes because they provide a stable, interpretable, and versionable surface for steering scientific workflows.
Specifically, using run cards as the orchestration boundary provides three key advantages:
\paragraph{Reproducibility and transparency.}
Run cards define complete, standalone configurations that can be version-controlled, archived, and executed independently of the agent. Each decision made by the LLM is therefore captured in a persistent artifact, enabling fully auditable workflows.

\paragraph{Compatibility with RAG and structured prompting.}
Because run cards follow stable syntax conventions, they are highly amenable to retrieval-augmented generation (RAG). The agent can be contextualized with validated examples, available parameter dictionaries, and relevant excerpts from official manuals, allowing it to generate syntactically correct and meaningful configurations without memorizing the entire documentation. This reduces the likelihood of malformed configurations and improves the reliability of free-form prompting. 
We leave a full implementation of RAG-based run-card generation to future iterations of the framework.

\paragraph{Human-approval checkpoints.}
Each generated run card constitutes a natural boundary at which the researcher may inspect, modify, or approve the configuration before execution. This preserves scientific judgment while still benefiting from automated orchestration.

\subsection{Context propagation and stateful execution}
\label{subsec:context}

\pkgname maintains a structured conversation state or \emph{context} that records tool outputs and relevant metadata. Later steps can reference this state directly, rather than relying on the agent to recall filenames or parse textual logs. For example, when a run card includes a user-defined parameter scan (e.g., multiple values of a mass or coupling), the \textsc{MadGraph} tool automatically detects the scan structure by inspecting the resulting run directories for characteristic naming patterns (\textit{e.g.,}~\texttt{run\_01}, \texttt{run\_02}, \ldots) and parsing the scan summary files generated by \textsc{MadGraph}.
Instead of returning only LHE file paths, the tool produces structured metadata:
\medskip
\begin{graybox}
\begin{verbatim}
{ "scan_detected": true,
  "n_runs": 3,
  "runs": [
        {"run_id": "run_01", "lhe_file": "...", ...},
        {"run_id": "run_02", "lhe_file": "...", ...},
        {"run_id": "run_03", "lhe_file": "...", ...}
]}
\end{verbatim}
\end{graybox}
associating each scan point with its corresponding parameter values, file locations, and computed cross sections.
This metadata is written into the conversation state, allowing subsequent steps, such as \textsc{Pythia}, to access the scan information directly without requiring the agent to interpret directory names or rely on textual recall.

Additionally, tools automatically detect and substitute placeholder values in run cards at execution time.
For instance, a \textsc{MadGraph} command card may contain \texttt{import model [[UFO\_PATH]]}, or a \textsc{Pythia} configuration may specify \texttt{Beams:LHEF = [[LHEF\_PATH]]}.
When invoked, the corresponding tool identifies these bracketed placeholders and replaces them with the actual file paths supplied as arguments (\textit{e.g.,}~the UFO directory from \textsc{FeynRules} output or the LHE file from a specific scan point):
\medskip
\begin{graybox}
\begin{verbatim}
# Template run card
Beams:LHEF = [[LHEF_PATH]]

# Automatically becomes
Beams:LHEF = data/mg_run001/.../run_01/events.lhe.gz
\end{verbatim}
\end{graybox}
This eliminates the need for agents to manually edit configuration files or construct fragile string-replacement logic, enabling run cards to serve as reusable templates across different workflow invocations while maintaining full traceability of substituted values in the execution log.

Long multi-stage workflows raise the question of \emph{context rot}, the degradation of model performance as the active context grows. \pkgname addresses this at several levels. Conversation-context management is the responsibility of the orchestration harness (\textsc{Orchestral AI}), which maintains the evolving context and supports pluggable compaction strategies, for example retaining a sliding window of recent turns as the context grows. \pkgname itself is built to be cognizant of this budget, with tool descriptions and outputs kept deliberately concise so that each step contributes a compact structured summary rather than raw logs. Bulky artifacts such as event records, generator logs, and scan summaries are written to the sandbox filesystem and referenced by path rather than held in-context, while the full sequence of tool invocations, arguments, and structured outputs is persisted to disk as an append-only record that gives a markdown-/log-style audit trail re-readable on demand. As the horizon of an automated task lengthens, further techniques become attractive, including version-control checkpoints such as automatic git commits of the sandbox state at phase boundaries for rollback and diffing, selective retrieval of only the relevant prior state back into context, and delegation of self-contained subtasks to sub-agents with their own context.

Together, schema-validated tools, stateful context propagation, and run-card-driven orchestration provide a robust foundation for LLM-assisted workflows.
These mechanisms ensure that multi-stage HEP analyses remain interpretable, reproducible, and aligned with established scientific practice.

\subsection{Orchestration versus scripting}

Before turning to a full workflow demonstration, it is useful to contrast the orchestration model with traditional scripting approaches commonly used to automate HEP simulations.

While scripting remains the standard approach for automating HEP workflows, it imposes important structural constraints. 
A bash script or Python pipeline can sequentially invoke \textsc{FeynRules}, \textsc{MadGraph}, \textsc{Pythia}, and downstream analysis tools. 
However, scripting enforces static control flow: parameters, file paths, and branching logic are embedded directly in code. 
As a result, adapting a workflow such as changing scan configurations, modifying analysis logic, or responding to tool failures requires manual intervention and re-editing of implementation details. 
Moreover, scripts provide no formal mechanism for representing workflow intent or ensuring consistency of metadata across heterogeneous tools.
These structural limitations can be alleviated through the introduction of an explicit orchestration layer with the following properties:

\paragraph{Adaptive workflow construction.}
Workflows are specified at the level of physics intent rather than program control flow. 
The LLM constructs executable plans by selecting appropriate tool invocations, propagating run-card metadata, and managing multi-stage execution.
This approach decouples workflow specification from its implementation, enabling dynamic reconfiguration without modifying underlying code.
This same mechanism directly addresses the consistency problem noted above. Quantities shared across run cards, e.g.\ UFO paths, LHE paths, scan points, and beam settings, are propagated automatically through placeholder substitution and context state (\cref{subsec:context}) rather than being re-entered by hand in each card, eliminating the most common source of inter-card inconsistency. Beyond this structural guarantee, because the agent can read every card, it can be instructed to perform explicit semantic cross-checks prior to execution, for example verifying that the \textsc{Pythia} beam configuration matches the \textsc{MadGraph} process or that the showering input references the intended scan point; a dedicated consistency-checking tool is a natural planned addition.

\paragraph{Structured error interpretation and recovery.}
Each tool exposes a schema-defined interface that includes structured failure modes. 
Instead of unstructured log output, tools return machine-interpretable error objects that the LLM can analyze to propose corrective actions or request clarification. This yields more resilient execution patterns and facilitates automated recovery strategies under human supervision.

\paragraph{Workflow transparency and provenance.}
Tool invocations and their outputs form an explicit execution trace, providing a machine-readable record of workflow decisions and intermediate artifacts. This provenance layer supports reproducibility, auditability, and collaboration, and is difficult to achieve in traditional scripts without significant custom infrastructure.

These capabilities introduce additional abstraction layers and rely on LLM reasoning quality, so for fixed or static pipelines, scripting may remain adequate. For exploratory analyses, parameter-dependent workflows, and multi-stage simulation studies, the orchestration layer provides stronger guarantees of adaptability, consistency, and traceability. Viewed together, these properties turn failures that would terminate a script into recoverable events, replace hand-written bookkeeping of paths and scan metadata, and prevent the silent format mismatches common to long shell pipelines. The resulting value proposition, for both the expert and non-expert user, is one of reliable and auditable assistance rather than raw speedup, as illustrated by the leptoquark workflow and quantified in \cref{sec:benchmarking}.

The following section returns to the leptoquark analysis introduced in \cref{sec:example} and illustrates these mechanisms in practice via a complete, multi-stage workflow orchestrated with \pkgname, showing how an agent constructs, executes, and analyzes a realistic HEP pipeline.

\section{Agent-orchestrated workflow}
\label{sec:example_full}

We now demonstrate how the \pkgname tool suite operates in practice by presenting an end-to-end execution of the leptoquark mass-scan workflow introduced in \cref{sec:example}. 
The goal is to illustrate how an agent, equipped with the orchestration mechanisms of \cref{sec:architecture}, constructs and executes a multi-stage HEP pipeline inside a pre-initialized sandbox while maintaining transparency, reproducibility, and human oversight.
The example was chosen deliberately because it exercises every architectural mechanism just described: schema-validated tool invocation, run-card-driven orchestration, scan-metadata propagation across stages, mixed tool-based and in-line execution, and structured error recovery. It is intended as a representative rather than exhaustive demonstration.

For concreteness, we retain the simplifying assumptions of the running example and consider a first-generation-philic scalar leptoquark with diagonal couplings to $(u,e)$, scanning over three benchmark mass points. 
The physics content of the example is unchanged; the focus here is the interaction between the agent and the orchestration layer.

\vspace{0.5em}
\noindent\textbf{Interaction modes}
Within the orchestration framework, there are several patterns of possible interaction between the user and the agent, each corresponding to a different level of planning autonomy. These modes share the same tool suite and execution environment but differ in how workflow intent is specified:

\begin{itemize}
  \item \textbf{To-do mode.}
  The user supplies a predefined, phase-structured to-do list (\texttt{todos.md}) describing the full workflow. 
  The agent executes tasks sequentially, updating task status and advancing through the prescribed stages with minimal additional input. This is closest to “programming” the workflow directly.

  \item \textbf{Planner mode.}
  The user provides a high-level objective. The agent autonomously constructs an explicit, step-by-step plan after inspecting the sandbox (model files, run cards, directory structure).
  The agent then executes this plan, invoking tools and verifying outputs at each stage. 
  This mode highlights the planning and coordination capabilities enabled by schema-validated tools and run-card-driven configuration.

  \item \textbf{Explorer mode.}
  A general interactive mode in which the agent responds to user queries or instructions without producing a global plan. 
  This mode is useful for exploratory tasks, debugging, or incremental development of workflows.
\end{itemize}

Each conversation begins in a sandbox set up by \pkgname, initialized with the relevant model files and run cards provided by the user, and the agent is confined to this workspace by tool-level path checks. \textsc{Orchestral} additionally exposes pre-execution hooks through which a user can approve or reject individual tool calls, including the shell- and Python-execution tools, giving a human-in-the-loop check on agent actions \cite{orchestral-ai}. The tools are configured ahead of time to point at the relevant external HEP codes, with full installation and setup details documented in the repository. Stronger isolation, such as OS-level containerization or dedicated virtual machines, is a natural direction for hardening this sandbox in future work.
As the workflow proceeds, tools populate this workspace with intermediate and final output (UFO directories, LHE files, JSONL event streams, jets, reconstructed observables).
In \cref{subsec:example_convo}, the sandbox is initialized with the following directory structure:
\begin{orchestralsystemmessage}

\begin{minted}{text}
.
├── feynrules
│   └── models
│       └── S1_LQ_RR.fr
├── madgraph
│   └── cards
│       └── S1_LQ_RR_pp_lqlq_scan.mg5
└── pythia
    └── cards
        └── S1_LQ_RR_pp_ljlj.cmnd

7 directories, 3 files
\end{minted}
    \end{orchestralsystemmessage}
where \texttt{S1\_LQ\_RR.fr}, \texttt{S1\_LQ\_RR\_pp\_lqlq\_scan.mg5}, and \texttt{S1\_LQ\_RR\_pp\_ljlj.cmnd} are assumed to be syntactically valid model files/run cards for \textsc{FeynRules}, \textsc{MadGraph}, and \textsc{Pythia}, respectively.
In to-do mode conversations, a user-defined todo-list is also copied into the sandbox.

\subsection{Example conversation}
\label{subsec:example_convo}

Below we provide an annotated transcript of a ``planning'' conversation with a GPT-OSS-120B \cite{openai2025gptoss120bgptoss20bmodel} agent. 
Throughout this example, the agent operates over pre-existing, syntactically valid \textsc{FeynRules}, \textsc{MadGraph}, and \textsc{Pythia} model files and run cards, managing their invocation, parameterization, and file-level dependencies rather than constructing the run cards themselves.
Note that some tool invocations and outputs are simplified for pedagogical clarity and brevity. 
Importantly, while this transcript shows only the JSON outputs returned by tools, the agent receives the full tool docstring (describing inputs, outputs, and scientific semantics) with each API call as described in \cref{sec:architecture}. 
For completeness, Appendix~\ref{app:tools} provides a brief summary of each tool and its primary use case, while the full tool docstrings are available in the public code repository.

\textbf{The system prompt.}
Conversational language models take as input a \emph{context} consisting of a structured list of messages with various roles such as \emph{user} or \emph{assistant}. 
Many frameworks also support specialized roles for \emph{tool} responses, and for \emph{system} messages. 
LLMs are generally fine tuned to strictly follow instructions given to them in these system messages. 
Most conversations begin with a ``system prompt'': a system message which shapes the LLM's behavior and can be used to define specialized context, goals, guidelines, formatting requirements, and operational constraints.
When using LLMs, one typically either needs to explicitly manage the context and append new messages, or use a framework which handles this. 
We implement our agents with \orchestralai, which automatically manages the context while allowing us to define our own custom system prompt.
The example below uses the following system prompt:

\begin{orchestralsystemmessage}
    You are a helpful assistant, expert high-energy-physicist, and professional computational scientist running on the Orchestral AI platform.
    Your job is to plan and execute beyond-the-Standard-Model high-energy-physics event generation and analysis workflows inside a pre-initialized sandbox containing directories with model files and run cards for event generation.
    Use tools ONLY when they provide a clear benefit, and NEVER fabricate physics or file contents.
    Markdown and LaTeX are allowed. Escape dollar signs except in math. No emojis.
    \vspace{0.05in}
    \begin{enumerate}
    \item \textbf{Workspace model}\\
    At the start of a session:
    \begin{enumerate}
        \item List the top-level directories and files.
        \item Inspect representative files with \texttt{readfile} to identify:\\
            -- which are model directories,\\
            -- which contain model files (\texttt{.fr}) or run cards (\texttt{.mg5}, \texttt{.cmnd}),\\
            -- collider configuration (beam type, energy),\\
            -- mass and coupling definitions.
    \end{enumerate}
    Treat these files as authoritative. Only modify them if the user explicitly asks, and otherwise
    create derived copies when changes are needed.
    \vspace{0.05in}
    \item \textbf{Workflow planning protocol}\\
    Your primary responsibility is structured planning followed by execution.\\
    \textit{Step 1: Clarify the goal}\\
    Summarize the user's objective in 1-3 sentences (e.g., event generation, validation, plotting).
    If already clear from context, restate without asking.

    \textit{Step 2: Write an explicit todo list}\\
    Create a to do list with concise steps that typically include:
    \begin{enumerate}
        \item Parameter selection and BSM model file generation.
        \item Event generation using the appropriate run card.
        \item Showering/hadronization.
        \item Analysis tasks (clustering, invariant masses, histograms).
        \item Plotting and summary outputs.
    \end{enumerate}

    \textit{Step 3: Execute step-by-step}\\
    For each planned step:
    \begin{enumerate}
        \item Invoke the correct tool with explicit arguments.
        \item Verify outputs (existence or small summary inspection).
        \item Provide a short progress message and proceed to the next step.
    \end{enumerate}
    \textit{\textbf{Work autonomously}} -- proceed through phases without waiting for the user.
    \vspace{0.05in}
    \item \textbf{Physics and workflow constraints}\\
    -- Do not alter collider settings unless instructed.\\
    -- Provide minimal run metadata after major tasks (inputs, seeds, versions).\\
    -- Give qualitative physics interpretations only; avoid strong claims without explicit
    statistical instruction.
\end{enumerate}

    \centerline{\vdots}
    \vspace{0.5em}

\end{orchestralsystemmessage}
\vspace{1.7in}

\noindent \textbf{The user prompt.}
Since LLM APIs operate as user-facing oracles, they require an explicit user prompt to generate a streamed response; the system prompt is transmitted implicitly and is not exposed to the API until the user submits the first prompt.

\begin{orchestralusermessage}
    Plan a Monte Carlo signal-validation analysis of the BSM scalar leptoquark defined by the supplied run cards. 
    Generate 10,000 events, select objects, and create lepton-jet invariant-mass distributions for each parameter point.
    Attempt to isolate and reconstruct the leptoquark resonance by applying reasonable lepton-jet separation and experimental cuts. 
    Compute and plot the resulting $m^{\text{min}}_{\text{LQ}}$ distribution. 
    Write a todo list first and then execute the todo list until completed.
\end{orchestralusermessage}

The agent responds by first recursively listing its current workspace:

\begin{orchestralagentmessage}

    \begin{orchestraltoolmessage}{RunCommand( \texttt{command=ls -R ., timeout=120000} )}
\begin{minted}{text}
Command: ls -R .
Return Code: 0
Output:
feynrules    madgraph    pythia

./feynrules:
models
./feynrules/models:
S1_LQ_RR.fr
./madgraph:
cards
./madgraph/cards:
S1_LQ_RR_pp_lqlq_scan.mg5
./pythia:
cards
./pythia/cards:
S1_LQ_RR_pp_ljlj.cmnd
\end{minted}
    \end{orchestraltoolmessage}

\end{orchestralagentmessage}
followed by the generation of a markdown todo list using the \texttt{TodoWriteTool}. The first item, marked with \texttt{[*]}, denotes that the task is in-progress.
\begin{orchestralagentmessage}

    \begin{orchestraltoolmessage}{ToDoWrite}
        \textbf{Todos:}\vspace{-0.2in}
\begin{minted}{text}
## Monte Carlo Signal-Validation Analysis for S1 Leptoquark

- [*] Generate UFO model from FeynRules file 
      `feynrules/models/S1_LQ_RR.fr`
- [ ] Generate parton-level events with MadGraph using card 
      `madgraph/cards/S1_LQ_RR_pp_lqlq_scan.mg5`
    - set `nevents = 10000`
    - keep all parameter points
- [ ] Shower and hadronize events with Pythia using card 
      `pythia/cards/S1_LQ_RR_pp_ljlj.cmnd`
    - `shower_lhe = True`
- [ ] Convert Pythia JSONL events to NumPy array
- [ ] Apply kinematic cuts to leptons and jets
    - leptons: pt>20 GeV, |eta|<2.5, PDG IDs 11,13 (and charge 
      conjugates)
    - jets: pt>30 GeV, |eta|<2.5
- [ ] Select hardest 2 leptons and hardest 2 jets per event
- [ ] Compute lepton-jet invariant masses for all pairings and retain
      the minimum per event (`m^min_{LQ}`)
- [ ] Plot the `m^min_{LQ}` distribution (histogram, 50 bins, 
      appropriate range) and save as `mmin_LQ.png`
- [ ] Summarize results (number of events, cross-section, plot 
      location)
\end{minted}
    \end{orchestraltoolmessage}
\end{orchestralagentmessage}
The todo list then proceeds through the event-generation tool chain, beginning with UFO model generation in \textsc{FeynRules}, followed sequentially by parton-level event generation with \textsc{MadGraph} and showering and hadronization with \textsc{Pythia}:
\begin{orchestralagentmessage}
    \begin{orchestraltoolmessageml}{FeynRulesToUFO( \texttt{model\_path=feynrules/models/S1\_LQ\_RR.fr,} \\\hphantom{FeynRulesToUFO(}
        \texttt{output\_dir=feynrules/models/S1\_LQ\_RR\_UFO,}\\\hphantom{FeynRulesToUFO(}
        \texttt{timeout\_sec=1800} )
    }

    \textbf{Output logs:} \vspace{-0.2in}
        \begin{minted}[escapeinside=||]{text}
{
    "schema_version": "tool-1.0",
    "ok": true,
    "model_path": "feynrules/models/S1_LQ_RR.fr",
    "output_dir": "feynrules/models/S1_LQ_RR_UFO",
    "wolframscript_path": "<path/to/wolframscript>",
    "feynrules_path": "<path/to/feynrules>",
    "logs": {
    "stdout": "feynrules/models/S1_LQ_RR_UFO/_logs/wolframscript_stdout.log",
    "stderr": "feynrules/models/S1_LQ_RR_UFO/_logs/wolframscript_stderr.log"
    },
    "files_created": [
    "couplings.py",
    "decays.py",
    "lorentz.py",
    "parameters.py",
    "particles.py",
    "propagators.py",
    "vertices.py",
         |$\vdots$|
    ]
}
        \end{minted}
    \end{orchestraltoolmessageml}

    \begin{orchestraltoolmessage}{ReadFile( \texttt{limit=200, path=madgraph/cards/S1\_LQ\_RR\_pp\_lqlq.mg5} )}
    \begin{minted}[escapeinside=||]{text}
File: madgraph/cards/S1_LQ_RR_pp_lqlq_scan.mg5 (35 lines)
 1→ set automatic_html_opening False
 2→ 
 3→ # Load the Standard Model UFO (built in)
 4→ import model [[UFO_PATH]]
 5→ 
 6→ # Define beam contents (proton = quarks + gluon)
 7→ define p = g u c d s b u~ c~ d~ s~ b~
 8→ 
10→ # Define a simple process:
11→ generate p p > S1 S1~, (S1 > e- u), (S1~ > e+  u~)
12→                     |$\vdots$|
    \end{minted}
    \end{orchestraltoolmessage}

    \begin{orchestraltoolmessageml}{MadgraphFromRunCard(\texttt{run\_card=<path/to>/S1\_LQ\_RR\_pp\_lqlq.mg5, nevents=10000} \\\hphantom{MadgraphFromRunCard(}\texttt{data\_dir=data/mg\_run001, output=S1\_LQ\_scan,}\\\hphantom{MadgraphFromRunCard(}\texttt{ufo\_path=feynrules/models/S1\_LQ\_RR\_UFO} )}
        \textbf{Output logs:} \vspace{-0.2in}
\begin{minted}[escapeinside=||]{text}
{"status":"ok",
"scan_detected":true,"data_dir":"data/mg_run001",
"lhe_file":"data/mg_run001/S1_LQ_scan/Events
            /run_01/unweighted_events.lhe.gz",
"manifest_json":"data/mg_run001/manifest.json",
"log_file":"data/mg_run001/mg5_run.log",
"scan_summary_file":"data/mg_run001/S1_LQ_scan/Events
                     /scan_run_0[1-3].txt",
"n_runs":3,
"runs":[
        {"run_id":"run_01",
         "lhe_file":"data/mg_run001/S1_LQ_scan/Events/
                     run_01/unweighted_events.lhe.gz",
         "scan_params":{"mass#9000005":1000.0},
         "cross_section":0.1016761,
         "cross_section_error":0.0001923201},
        {"run_id":"run_02",
         "lhe_file":"data/mg_run001/S1_LQ_scan/Events/
                     run_02/unweighted_events.lhe.gz",
         "scan_params":{"mass#9000005":1500.0},
         "cross_section":0.008729035,
         "cross_section_error":1.541146e-05},
                          |$\vdots$|
        ]
    }
\end{minted}
    \end{orchestraltoolmessageml}

    \begin{orchestraltoolmessage}{ReadFile( \texttt{limit=200, path=pythia/cards/S1\_LQ\_RR\_pp\_ljlj.cmnd} )}
\begin{minted}{text}
File: pythia/cards/S1_LQ_RR_pp_ljlj.cmnd (11 lines)
 1→ #------------------------------------------------------------
 2→ # Example: pp_S1S1d_ljlj.cmnd
 3→ # Minimal Pythia8 run card to shower LHE events from MadGraph
 4→ #------------------------------------------------------------
 5→ 
 6→ # Beam and input setup
 7→ Beams:frameType = 4
 8→ Beams:LHEF      = [[LHEF_PATH]]
 9→ 
10→ # Quiet output
11→ Print:quiet = on
\end{minted}
    \end{orchestraltoolmessage}

    \begin{orchestraltoolmessageml}{PythiaFromRunCard(\texttt{ cmnd\_path="pythia/cards/S1\_LQ\_RR\_pp\_ljlj.cmnd",}\\\hphantom{PythiaFromRunCard(} 
    \texttt{data\_dir="data/pythia\_run01", finals\_only="True",}\\\hphantom{PythiaFromRunCard(} \texttt{lhe\_path="<path/to>/run\_01/unweighted\_events.lhe.gz", ... )}}
        \textbf{Output logs:}\vspace{-0.2in}

\begin{minted}{text}
{"status":"ok",
"data_dir":"data/pythia_run01",
"events_jsonl":"data/pythia_run01/events.jsonl",
"manifest_json":"data/pythia_run01/manifest.json",
"accepted":10000,
"failed":0}
\end{minted}
    \end{orchestraltoolmessageml}

    \begin{orchestraltoolmessage}{PythiaFromRunCard( $\cdots$ )}
        Same as above with \texttt{run\_01} $\to$ \texttt{run\_02}
    \end{orchestraltoolmessage}

    \begin{orchestraltoolmessage}{PythiaFromRunCard( $\cdots$ )}
        Same as above with \texttt{run\_02} $\to$ \texttt{run\_03}
    \end{orchestraltoolmessage}

    \begin{orchestraltoolmessageml}{JetClusterSlowJet( \texttt{R=0.4, algorithm=antikt, cluster\_all=True, etamax=5,}\\\hphantom{JetClusterSlowJet(}
         \texttt{jsonl\_path=data/pythia\_run01/events.jsonl, mass\_option=1,}\\\hphantom{JetClusterSlowJet(} \texttt{output\_path=data/pythia\_run01/jets.jsonl, ptmin=0} )}
\begin{minted}{text}
{"status":"ok",
"n_events":10000,
"output_file":"data/pythia_run01/jets.jsonl"}
\end{minted}
    \end{orchestraltoolmessageml}

    \begin{orchestraltoolmessage}{JetClusterSlowJet( $\cdots$ )}
        Same as above with \texttt{run\_01} $\to$ \texttt{run\_02}
    \end{orchestraltoolmessage}

    \begin{orchestraltoolmessage}{ToDoWrite}
        \textbf{Todos:}\vspace{-0.2in}

\begin{minted}{text}
## Monte Carlo Signal‑Validation Analysis for S1 Leptoquark

- [x] Generate UFO model from FeynRules file 
      `feynrules/models/S1_LQ_RR.fr`
- [x] Generate parton‑level events with MadGraph using card 
      `madgraph/cards/S1_LQ_RR_pp_lqlq_scan.mg5`
    - set `nevents = 10000`
    - keep all parameter points
- [x] Shower and hadronize events with Pythia using card 
      `pythia/cards/S1_LQ_RR_pp_ljlj.cmnd`
    - `shower_lhe = True`
- [x] Cluster jets for each run
    - run01 jets `data/pythia_run01/jets.jsonl`
    - run02 jets `data/pythia_run02/jets.jsonl`
    - run03 jets `data/pythia_run03/jets.jsonl`
- [*] Select hardest objects and compute m^min_{LQ}
    - select hardest 2 leptons (pdg 11, -11, 13, -13) per event
    - select hardest 2 jets per event
    - compute invariant masses for all lepton-jet pairings, 
      keep minimum per event
    - plot histogram of m^min_{LQ} (50 bins) and save as 
      `mmin_LQ.png`
- [ ] Summarize results (event counts, cross-sections, 
      plot location)
\end{minted}
    \end{orchestraltoolmessage}

    \begin{orchestraltoolmessageml}{FilterByPDGID(  \texttt{input\_path=data/pythia\_run01/events.jsonl,}\\
        \hphantom{FilterByPDGID(} \texttt{output\_path=data/pythia\_run01/leptons\_all.jsonl,}\\\hphantom{FilterByPDGID(}
        \texttt{pdgids=[11, -11, 13, -13]} )}
\begin{minted}{text}
{"status":"ok",
"input_events":10000,
"particles_before":5241292,
"particles_after":54846,
"kept_pdgids":[11,-11,13,-13],
"output_path":"data/pythia_run01/leptons_all.jsonl"}
\end{minted}
    \end{orchestraltoolmessageml}

    \begin{orchestraltoolmessageml}{GetHardestN( \texttt{input\_path=data/pythia\_run01/leptons\_a.ll.jsonl,}\\
        \hphantom{GetHardestN(}
         \texttt{n\_hardest=2,}\\
        \hphantom{GetHardestN(} \texttt{output\_path=data/pythia\_run01/hardest2\_leptons.jsonl} )}
\begin{minted}{text}
{"status":"ok",
"input_events":10000,
"n_hardest":2,
"output_path":"data/pythia_run01/hardest2_leptons.jsonl"}
\end{minted}
    \end{orchestraltoolmessageml}

    \begin{orchestraltoolmessageml}{GetHardestNJets(  \texttt{input\_path=data/pythia\_run01/jets.jsonl,}\\\hphantom{GetHardestNJets(} 
        \texttt{n\_hardest=2,}\\\hphantom{GetHardestNJets(}
        \texttt{output\_path=data/pythia\_run01/hardest2\_jets.jsonl} )}
\begin{minted}{text}
{"status":"ok",
"input_events":10000,
"n_hardest":2,
"output_path":"data/pythia_run01/hardest2_jets.jsonl"}
\end{minted}
    \end{orchestraltoolmessageml}

    \begin{orchestraltoolmessageml}{ResonanceReconstruction( \texttt{min\_delta\_r=0.4, output\_prefix=outputs/run01\_LQ,}\\\hphantom{ResonanceReconstruction(} \texttt{particle\_arrays=[<path/to>/hardest2\_leptons.jsonl, ...]}\\\hphantom{ResonanceReconstruction(}
    \texttt{template=two\_body\_symmetric} )}
\begin{minted}{text}
{"status":"ok",
"template":"two_body_symmetric",
"n_arrays":2, ...
"n_events_analyzed":10000,
"n_events_successful":9830, ...,
"observables":[...],
"histograms":[{"observable":"m1",
"bins":[...], "counts":[...]},
{"observable":"m2",
"bins":[...], "counts":[...]},
{"observable":"m_min",
"bins":[...], "counts":[...]},
{"observable":"m_max",
"bins":[...],
"counts":[...]}],
"data_paths":{...,
"m_min":"outputs/run01_LQ_m_min.npy",
"m_max":"outputs/run01_LQ_m_max.npy"}}
\end{minted}
    \end{orchestraltoolmessageml}

    \begin{orchestraltoolmessage}{ReadFile( \texttt{limit=200, path=data/mg\_run001/manifest.json} )}
\begin{minted}[escapeinside=||]{text}
File: data/mg_run001/manifest.json (20 lines)
                                    |$\vdots$|
\end{minted}
    \end{orchestraltoolmessage}

    \begin{orchestraltoolmessage}{FilterByPDGID( $\cdots$ )
        }
    Same as \textsf{FilterByPDGID} above with \texttt{run\_01} $\to$ \texttt{run\_02}
    \end{orchestraltoolmessage}

    \begin{orchestraltoolmessage}{FilterByPDGID( $\cdots$ )
        }
    Same as \textsf{FilterByPDGID} above with \texttt{run\_02} $\to$ \texttt{run\_03}
    \end{orchestraltoolmessage}

    \begin{orchestraltoolmessage}{GetHardestN( $\cdots$ )}
    Same as \textsf{GetHardestN} above with \texttt{run\_01} $\to$ \texttt{run\_02}
    \end{orchestraltoolmessage}

    \begin{orchestraltoolmessage}{GetHardestNJets( $\cdots$ )}
    Same as \textsf{GetHardestNJets} above with \texttt{run\_01} $\to$ \texttt{run\_02}
    \end{orchestraltoolmessage}

    \begin{orchestraltoolmessageml}{ResonanceReconstruction( \texttt{min\_delta\_r=0.4, output\_prefix=outputs/run02\_LQ,}\\\hphantom{ResonanceReconstruction(} \texttt{particle\_arrays=['data/pythia\_run02/hardest...,}\\\hphantom{ResonanceReconstruction(}
    \texttt{template=two\_body\_symmetric}}
\begin{minted}[escapeinside=||]{text}
|\textcolor{red}{Error: Processing Error}|
- Reason: Array 1 not found: data/pythia_run02/hardest2_jets.jsonl
\end{minted}
    \end{orchestraltoolmessageml}

    \begin{orchestraltoolmessageml}{GetHardestNJets(  \texttt{input\_path=data/pythia\_run03/jets.jsonl,}\\\hphantom{GetHardestNJets(} 
            \texttt{n\_hardest=2,}\\\hphantom{GetHardestNJets(}
            \texttt{output\_path=data/pythia\_run03/hardest2\_...}}
\begin{minted}[escapeinside=||]{text}
|\textcolor{red}{Error: File Not Found}|
- Reason: Input file not found: data/pythia_run03/jets.jsonl
\end{minted}
    \end{orchestraltoolmessageml}

    \begin{orchestraltoolmessage}{JetClusterSlowJet( $\cdots$ )}
        Same as above with \texttt{run\_02} $\to$ \texttt{run\_03}
    \end{orchestraltoolmessage}

    \begin{orchestraltoolmessage}{GetHardestN( $\cdots$ )}
    Same as \textsf{GetHardestN} above with \texttt{run\_02} $\to$ \texttt{run\_03}
    \end{orchestraltoolmessage}

    \begin{orchestraltoolmessage}{GetHardestNJets( $\cdots$ )}
    Same as \textsf{GetHardestNJets} above with \texttt{run\_02} $\to$ \texttt{run\_03}
    \end{orchestraltoolmessage}

    \begin{orchestraltoolmessage}{ResonanceReconstruction( $\cdots$ )}
    Retry of \texttt{run\_02}
    \end{orchestraltoolmessage}

\begin{orchestraltoolmessage}{ResonanceReconstruction( $\cdots$ )}
    Same as \textsf{ResonanceReconstruction} above with \texttt{run\_02} $\to$ \texttt{run\_03}
    \end{orchestraltoolmessage}

\end{orchestralagentmessage}

In the above tool chain, note the automated error-handling behavior that occurs when the agent incorrectly assumes that jet clustering had already been completed for all scan points. 
The tool execution layer returns a structured error, enabling the agent to identify and correct the missing step rather than encountering an opaque runtime failure that would typically terminate a conventionally scripted workflow. 
The agent responds by re-invoking the appropriate jet-clustering step for the affected run before successfully resuming resonance reconstruction.
Finally, the agent finishes execution by generating the requested plot and summarizing the results.
This final plotting step illustrates free-form, in-line code generation rather than tool invocation (see \cref{subsec:goodtool}). Because the specific histogram is particular to this analysis and not known in advance, the agent writes and executes Python (via \texttt{RunPython}) to produce the $m^{\min}_{\text{LQ}}$ histogram rather than calling a dedicated plotting tool.

\begin{orchestralagentmessage}

    \begin{orchestraltoolmessage}{RunPython}
        \textbf{Code:}\vspace{-0.2in}

        \begin{minted}{python}
    import subprocess, os, sys, json, numpy as np, matplotlib.pyplot as plt

    # Ensure matplotlib backend for headless
    import matplotlib
    matplotlib.use('Agg')

    paths = {
        '1000 GeV': 'outputs/run01_LQ_m_min.npy',
        '1500 GeV': 'outputs/run02_LQ_m_min.npy',
        '2000 GeV': 'outputs/run03_LQ_m_min.npy'
    }

    data = {label: np.load(p) for label,p in paths.items()}

    plt.figure(figsize=(8,6))
    for label, arr in data.items():
        plt.hist(arr, bins=50, histtype='step', label=label)
    plt.xlabel('m_{min}^{LQ} [GeV]')
    plt.ylabel('Events')
    plt.title('Minimum lepton‑jet invariant mass per event')
    plt.legend()
    plt.grid(True)
    plt.tight_layout()
    plt.savefig('mmin_LQ.png')
    print('Plot saved to mmin_LQ.png')

        \end{minted}

        \textbf{Output:}\vspace{-0.2in}

        \begin{minted}{text}
    {"stdout": "Plot saved to mmin_LQ.png", "stderr": "", "return_code": 0}
        \end{minted}
    \end{orchestraltoolmessage}

    \vspace{0.1in}

    \textbf{Completed workflow:}

    \begin{center}
    \begin{tabularx}{\textwidth}{p{1cm} X X}
    \hline
    \textbf{Step} & \textbf{Description} & \textbf{Outcome} \\
    \hline
    1 & Generate UFO model from \texttt{feynrules/models/S1\_LQ\_RR.fr} & UFO created in \texttt{feynrules/models/S1\_LQ\_RR\_UFO} \\
    2 & Generate parton-level events with MadGraph (3 scan points: 1000 GeV, 1500 GeV, 2000 GeV) & 10,000 LHE events per point (\texttt{data/mg\_run001/...}) \\
    3 & Shower \& hadronize with Pythia8 (\texttt{pythia/cards/S1\_LQ\_RR\_pp\_ljlj.cmnd}) & 10,000 final-state events per point (\texttt{data/pythia\_run0X/events.jsonl}) \\
    4 & Jet clustering (anti-$k_t$ $R=0.4$) & Jets stored in \texttt{data/pythia\_run0X/jets.jsonl} \\
    5 & Lepton selection (electrons $\pm 11$, muons $\pm 13$) & \texttt{leptons\_all.jsonl} for each run \\
    6 & Keep hardest 2 leptons \& hardest 2 jets per event & \texttt{hardest2\_leptons.jsonl}, \texttt{hardest2\_jets.jsonl} \\
    7 & Reconstruct leptoquark candidates (two-body symmetric template, $\Delta R > 0.4$) & Minimum lepton-jet invariant mass $m_{\text{min}}$ saved as NumPy arrays \texttt{outputs/run0X\_LQ\_m\_min.npy} \\
    8 & Plot the distribution of $m^{\text{min}}_{\text{LQ}}$ for the three mass points & Histogram saved as \texttt{mmin\_LQ.png} \\
    9 & Summarize results & See below \\
    \hline
    \end{tabularx}
    \end{center}
    \vspace{0.1in}
    
    \textbf{Plot}

    \vspace{0.1in}
    The file \texttt{mmin\_LQ.png} contains over-laid histograms (50 bins) of the minimum lepton-jet invariant mass per event for the three leptoquark mass hypotheses (1000 GeV, 1500 GeV, 2000 GeV). The peaks shift upward with the leptoquark mass, demonstrating successful reconstruction.

    \vspace{0.1in}
    \textbf{Cross-section \& event counts}\\

    \begin{center}
    \begin{tabular}{cccc}
    \hline
    \textbf{Mass point} & \textbf{Scan parameter (GeV)} & \textbf{Cross-section (mb)} & \textbf{Events generated} \\
    \hline
    Run 01 & 1000 & $\cdots$ & 10,000 \\
    Run 02 & 1500 & $\cdots$ & 10,000 \\
    Run 03 & 2000 & $\cdots$ & 10,000 \\
    \hline
    \end{tabular}
    \end{center}

    \vspace{0.1in}
    \textbf{Key output files}
    \vspace{0.1in}
    \begin{itemize}
      \item UFO model: \texttt{feynrules/models/S1\_LQ\_RR\_UFO/}  
      \item MadGraph LHE files: \texttt{data/mg\_run001/S1\_LQ\_scan/Events/\\
      \quad \quad run\_XX/unweighted\_events.lhe.gz}  
      \item Pythia events (JSONL): \texttt{data/pythia\_run0X/events.jsonl}  
      \item Jet collections: \texttt{data/pythia\_run0X/jets.jsonl}  
      \item Selected objects: \texttt{data/pythia\_run0X/hardest2\_leptons.jsonl}, \\\texttt{data/pythia\_run0X/hardest2\_jets.jsonl}  
      \item Resonance observables: \texttt{outputs/run0X\_LQ\_m\_min.npy} (and m1, m2, m\_max)  
      \item Final plot: \texttt{mmin\_LQ.png}  
    \end{itemize}

    \vspace{0.1in}
    \textbf{Todo list (all tasks completed)}\vspace{-0.1in}

    \begin{minted}[escapeinside=||]{markdown}
    ## Monte Carlo Signal-Validation Analysis for S1 Leptoquark
                                            |\vdots|
    - [x] Select hardest objects and compute |$m^{\min}_{\text{LQ}}$|
        - select hardest 2 leptons (pdg 11, -11, 13, -13) per event
        - select hardest 2 jets per event
        - compute invariant masses for all lepton-jet pairings, keep minimum per 
          event
        - plot histogram of |$m^{\min}_{\text{LQ}}$| (50 bins) and save as
          `mmin_LQ.png`
    - [x] Summarize results (event counts, cross-sections, plot location)
    \end{minted}

    All steps have been executed and the requested analysis is finished. Let me know if you need any further inspection (e.g., individual event listings, alternative cuts, or additional plots).

\end{orchestralagentmessage}

\begin{figure}[t!]
    \centering
    \includegraphics[width=0.85\textwidth]{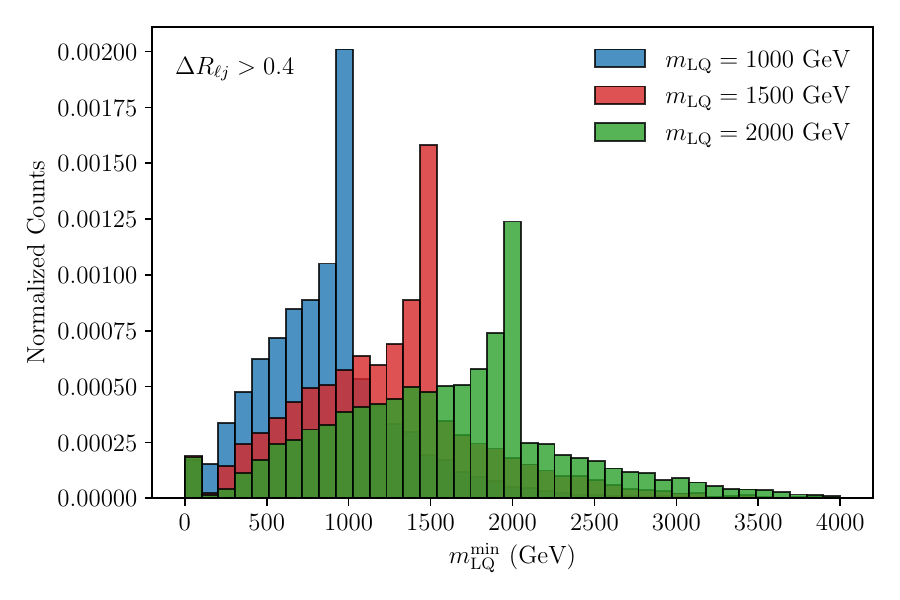}
    \caption{Reconstructed minimum leptoquark mass $m^{\text{min}}_{\text{LQ}}$ distributions for three benchmark scalar leptoquark mass points: $m_{S_1} = 1.0, 1.5, 2.0~\text{TeV}$.
    The agent-orchestrated the full simulation chain (\textsc{FeynRules}, \textsc{MadGraph}, \textsc{Pythia}, jet clustering, and invariant mass reconstruction) for each scan point, producing resonance peaks centered at the corresponding generated masses.
    Each distribution contains 10,000 simulated events with symmetric two-body decay reconstruction $(\ell j)(\ell j)$ and $\Delta R > 0.4$ separation cuts.}\label{fig:resonance_reconstruction}
\end{figure}

The complete leptoquark mass-scan example illustrates how these components function together in practice. 
Beginning from a pre-initialized sandbox, the agent autonomously identified available model and run-card files, generated a UFO model with \textsc{FeynRules}, executed a parameter scan with \textsc{MadGraph}, and showered each mass point with \textsc{Pythia}. 
It then carried out jet clustering, object selection, and resonance reconstruction using schema-validated analysis tools. 
A notable feature of the workflow is the agent's ability to propagate scan metadata across stages without manual bookkeeping, recover from intermediate failures, and maintain a structured execution trace throughout without input from the user. 
The resulting minimum lepton-jet invariant mass distributions for the three signal benchmarks are shown in \cref{fig:resonance_reconstruction}, demonstrating that the orchestrated workflow successfully reconstructs the expected resonance peaks and produces reproducible, analysis-ready outputs.

\paragraph{Range and limitations of the present tool suite.}
The tool suite supports a broad class of simulation-based workflows that reuse the same tool chain with different run cards, for example the dominant SM background channels to the present signal (Drell--Yan$\,+\,$jets, $t\bar t$, diboson), other BSM signals, and differential cross-section or cut-flow studies. Operations for which the suite provides no dedicated tool, such as flavor tagging or missing-energy observables, can in many cases still be handled by free-form, in-line code generation (\cref{subsec:goodtool}), while genuinely new capabilities, such as detector-level simulation or a new external generator, are added by implementing a single schema-validated \texttt{BaseTool} subclass (\cref{app:tools}), after which the orchestration layer coordinates the new tool exactly as it does the existing ones. The leptoquark example illustrates how schema-validated domain tools can make a multi-stage HEP workflow tractable for an open-weight backend such as GPT-OSS-120B. For stronger backends, the same interfaces remain useful as reproducibility, validation, and auditability infrastructure, even when the raw capability gain is smaller. A systematic, quantitative characterization of this range, across multiple workflows and LLM backends, is the subject of the dedicated benchmarking study previewed in \cref{sec:benchmarking} and developed in forthcoming work~\cite{HEPTAPOD:benchmarking}.

\section{Benchmarking}\label{sec:benchmarking}

To complement the qualitative demonstration of \cref{sec:example_full} with quantitative evidence, we evaluate the contribution of the orchestration layer for a fixed backend, GPT-OSS-120B \cite{openai2025gptoss120bgptoss20bmodel}.

\subsection{Task and initial conditions} The benchmark uses the leptoquark mass-scan task of \cref{sec:example_full}, with the user prompt shown in \cref{subsec:example_convo}. Each trial begins from the same full sandbox scaffold: the supplied \textsc{FeynRules} model file, the \textsc{MadGraph} and \textsc{Pythia} run cards, and the same three scan masses. The prompt specifies the physics goal but does not give the precise definition of $m_{\mathrm{LQ}}^{\min}$ given in \cref{eq:mLQmin}. 
A run \emph{succeeds} when the resulting distributions peak at all three injected scan masses. We compare two variants of this setup:
\begin{itemize}
  \item \textbf{Tool-assisted baseline:} the planner-mode agent (see \cref{subsec:example_convo}) with the full schema-validated tool suite of Appendix \ref{app:tools}.
  \item \textbf{Core-only recipes:} the same backend, user prompt, and sandbox scaffold, but with the HEPTAPOD domain tools removed. The agent has only \textsc{Orchestral}'s core shell, file, and Python capabilities. To make this comparison conservative, the system prompt supplies the relevant software paths and minimal verified invocation recipes for \textsc{FeynRules}$\to$UFO generation, \textsc{MadGraph} card execution, and \textsc{Pythia} showering through its Python bindings. This controls for tacit knowledge of how to launch the external programs, so the comparison isolates the contribution of the domain-tool harness beyond access to the underlying software.
\end{itemize}
We run each variant $N=10$ times and grade every run with the same content-based rubric (see App.~\ref{app:rubric}), whose stages mirror the pipeline above. Each stage is scored on \emph{evidence of the physics}, meaning an artifact of the right kind in any file format, rather than on a specific tool's output. This places the two variants on identical footing. The same stagewise rubric also defines the reach observable reported in Appendix \ref{app:rubric}, which measures partial progress on runs that do not complete the full workflow.

We summarize the performance of each variant with two metrics from the code-generation and agent-evaluation literature.  Writing $c$ for the number of successful runs (as defined above) among the $N$ trials, the capability metric pass@$k$~\cite{kulal2019spoc,chen2021evaluating} is the probability that at least one of $k$ independent attempts succeeds and the consistency metric pass$^k$~\cite{liu2024stablereasoning} is the probability that \emph{all} $k$ attempts succeed. Both are computed with the standard unbiased estimators
\begin{equation}
\text{pass@}k \;=\; 1 - \binom{N-c}{k}\Big/\binom{N}{k}, \qquad
\text{pass}^k \;=\; \binom{c}{k}\Big/\binom{N}{k}.
\label{eq:passk}
\end{equation}
pass@$1$ is the single-shot success rate, and the growth of pass@$k$ with $k$ measures the value of repeated sampling. The pass$^k$ metric instead asks whether every sampled run succeeds, making it a useful proxy for reliability in workflows where any single failure requires human intervention.

\subsection{Results} 
\begin{figure}[t]
\centering
\includegraphics[width=0.75\linewidth]{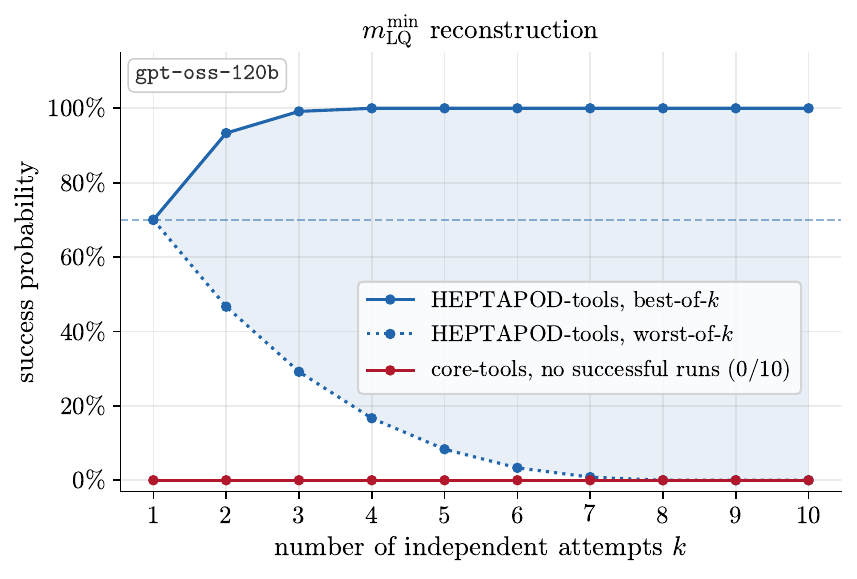}
\caption{Reliability vs.\ repetition on the leptoquark workflow (GPT-OSS-120B, $N=10$). The figure shows the probability of a fully successful run as a function of the number of independent attempts $k$, for the tool-assisted (blue) and core-only (red) variants. For the tool-assisted variant, the solid curve shows pass@$k$ (at least one of $k$ attempts succeeds), the dotted curve shows pass$^k$ (all $k$ attempts succeed), and the dashed line marks the single-shot mean; with domain tools, pass@$k$ reaches unity within a few attempts. The core-only variant never completes the workflow, so its pass@$k$ and pass$^k$ are identically zero at every $k$ and coincide in the single solid line at $0\%$ ($0/10$ runs).}
\label{fig:passk}
\end{figure}

\Cref{fig:passk} summarizes the two variants. With the domain-tool suite, the agent completes the full simulation pipeline in every run and reconstructs the resonance in $7/10$ single attempts ($\text{pass@}1=0.70$). The corresponding pass@$k$ reaches unity within a few attempts. In the core-only variant, even with the conservative invocation recipes, the same model never completes the workflow ($\text{pass@}1=\text{pass@}k=0$). Two distinct mechanisms account for the gap. First, driving the programs by hand requires the model to emit large tool-call payloads, including entire scripts and run cards passed as arguments. GPT-OSS-120B frequently fails to serialize these payloads as valid JSON. The runner treats such malformed tool-call JSON as a format crash and, up to two times, resumes the same session with the parser error and asks the model to re-issue the offending call as literal JSON. In $9/10$ core-only runs, at least one such decoding failure occurs, and roughly half of these failures remain fatal after the allowed retries. Second, on the runs that do execute the machinery, the model reconstructs the observable incorrectly. The $m_{\mathrm{LQ}}^{\min}$ distributions are empty or collapse to zero rather than peaking at the injected masses. As a result, $0/10$ core-only runs yield a correct reconstruction versus $7/10$ with domain tools. The orchestration layer thus contributes in two ways: (1) it reduces the serialization burden by exposing schema-validated inputs, and (2) it prepares structured, physics-level intermediate artifacts that make the final reconstruction a smaller and more checkable coding task. The HEPTAPOD tools turn a workflow that is \emph{never} completed autonomously with core tools alone into one that succeeds in the majority of attempts.

\vspace{-0.1in}
\section{Conclusions}\label{sec:conclusions}

In this work, we introduced \pkgname, an orchestration framework that integrates large language models into HEP workflows.
By treating scientific software as a collection of formally specified operations, \pkgname enables LLMs to coordinate domain-specific codes within transparent, human-supervised workflows.
We emphasize that the contribution of this paper lies in the orchestration \emph{infrastructure} itself, including schema-validated tools, run-card-driven coordination, stateful context propagation, structured error handling, and the LLM-legible \texttt{evtjsonl} format, together with a complete, reproducible end-to-end demonstration.
The complete leptoquark analysis demonstrated how these components combine to form a reproducible multi-stage pipeline spanning model definition, event generation, showering, reconstruction, and downstream analysis, turning GPT-OSS-120B into a capable phenomenological assistant for this workflow.

\pkgname{} complements the \orchestralai engine with HEP-specific abstractions, including schema-validated tools, run-card–driven configuration, and LLM-compatible event data structures. By treating run cards, intermediate artifacts, and structured event representations as first-class objects within a persistent orchestration loop, \pkgname{} provides a stable and auditable interface linking symbolic theory inputs, numerical configuration choices, and multi-stage simulation workflows. This design enables the seamless construction and execution of end-to-end HEP pipelines while preserving transparency, provenance, and human oversight throughout.

Looking forward, increasingly agentic scientific systems will likely play a growing role in HEP research. 
With reliable tool interfaces and structured context, LLM-driven agents could assist in tasks such as generator tuning, parameter-space exploration, automated background modeling, or adaptive analysis strategies that respond to intermediate results. 
More capable multi-agent configurations may support iterative model refinement, where one agent proposes theoretical extensions, another evaluates phenomenological viability, and a third orchestrates simulations to test consistency against data \cite{jiang2025agenticscimlcollaborativemultiagentsystems}. 
Such systems would not replace human judgment, but rather amplify it by automating the mechanical components of scientific computation while preserving interpretability and oversight.

The present architecture suggests several natural directions for future work.
First, run-card-based retrieval-augmented generation (RAG) offers a principled path toward more reliable configuration synthesis: curated collections of validated cards, parameter dictionaries, and tool-specific patterns could guide agents toward syntactically correct and semantically meaningful configurations without extensive supervision. 
A closely related extension is the automated extraction of phenomenological models from the literature. An agent could ingest a \texttt{hep-ph} manuscript, identify the relevant field content and interactions, and construct a corresponding \textsc{FeynRules} model file or an equivalent intermediate representation together with the other event-generator configurations, enabling end-to-end translation of theoretical models described in the literature to event-level simulations.
Second, extending the orchestration layer to closely related domains such as additional HEP event generators, \textsc{HepMC} \cite{Buckley:2019xhk} inspection and export, detector simulation, global analyses, and cross-generator validation represents the most direct progression from the capabilities demonstrated here. 
These additions would allow \pkgname{} to support full end-to-end phenomenological studies without altering the underlying orchestration philosophy.
Beyond these immediate extensions, \pkgname can also incorporate broader classes of HEP workflows, including neutrino event generators (e.g.~\textsc{Genie}\cite{Andreopoulos:2009rq}, \textsc{ACHILLES}\cite{Isaacson:2022cwh}), long-lived particle pipelines for beam-dump and intensity-frontier experiments, and detector-simulation interfaces for fixed-target and collider environments. 
Related opportunities include symbolic learning \cite{Alnuqaydan:2022ncd,Dong:2022trn}, model building \cite{Matchev:2024ash,Wojcik:2024lfy,Kawai:2024pws,Koay:2025bmu}, automating cross-generator consistency studies (e.g.~\textsc{Pythia}, \textsc{Sherpa}, \textsc{Herwig}), incorporating global-fit frameworks, or coordinating EFT-matching and parameter-inference tasks across multiple scales.
Third, progressively deeper forms of autonomy, such as agents that refine workflows in response to failures, track uncertainty across stages, or propose targeted simulations to resolve ambiguities, represent a promising evolution of the orchestration paradigm. In parallel, systematically logging agent decisions, tool invocations, and execution outcomes would enable the construction of domain-specific datasets for supervised fine-tuning and reinforcement learning, providing a path for improved agent performance through experience rather than prompt engineering alone. 

\pkgname provides the infrastructure required for these developments: a reliable, robust, auditable, and extensible orchestration layer through which future agentic systems can plan, execute, and refine HEP workflows. 
By linking symbolic theory to measurable data within a principled tool-driven environment, \pkgname offers a foundation for progressively more capable AI-assisted research in high-energy physics.

{\bf Acknowledgements.} The work of TM, AR and KM is supported in part by the Shelby Endowment for Distinguished Faculty at the University of Alabama. The work of SG is supported in part by the U.S. Department of Energy (DOE) under Awards No. DE-SC0012447 and No. DE-SC0026347. The work of KM is supported in part by the U.S. Department of Energy (DOE) under Award No. DE-SC0026347.
The work of TM is partially supported by Fermilab via Subcontract 725339. 
The work of AR and KM is partially supported by Fermilab via Subcontract 731293, in support of DOE Award No.\ DE-SCL0000090 ``HEP AmSC IDA Pilot: Knowledge Extraction'' and DOE Award No.\ DE-SCL0000152 ``USQCD AmSC Infrastructure Provision''.
PS is supported by the U.S. Department of Energy, Office of Science, Office of High Energy Physics QuantISED program under the grants ``HEP Machine Learning and Optimization Go Quantum'', Award Number 0000240323, and ``DOE QuantiSED Consortium QCCFP-QMLQCF'', Award Number DE-SC0019219.
This manuscript has been authored by Fermi Forward Discovery Group, LLC under Contract No. 89243024CSC000002 with the U.S. Department of Energy, Office of Science, Office of High Energy Physics.

{\bf Public code.} The open-source software developed for this project is publicly available at:
\begin{center}
{\url{https://github.com/tonymenzo/heptapod}}.
\end{center}

\newpage
\begin{appendix}

\section{Benchmark rubric}\label{app:rubric}

Each run in \cref{sec:benchmarking} is graded against a fixed, stagewise rubric that mirrors the leptoquark pipeline. A stage passes when a \emph{content-based} check finds the corresponding physics artifact, independent of which command or tool produced it. The accepted formats are deliberately broad. For example, showered events may be recorded as line-delimited JSON, \textsc{HepMC}, or a numeric array. This keeps the tool-assisted and core-only variants on identical footing. \Cref{tab:rubric} lists the stages, weights, and passing criteria.

\begin{table}[h!]
\centering
\begin{tabular}{llp{8.5cm}}
\hline
\textbf{Stage} & \textbf{Wt.} & \textbf{Passes when the run has produced\ldots} \\
\hline\\[-0.1in]
UFO model        & $0.10$ & a valid UFO model directory (particle, parameter, vertex, and coupling modules). \\
Parton events    & $0.15$ & $\ge 100$ parton-level events (LHE) across the parameter scan. \\
Showering        & $0.10$ & $\ge 100$ per-event showered-particle records. \\
Jet clustering   & $0.10$ & $\ge 100$ per-event clustered-jet records. \\
Reconstruction   & $0.05$ & $\ge 50$ reconstructed $m_{\mathrm{LQ}}^{\min}$ values. \\
Correct mass     & $0.50$ & an $m_{\mathrm{LQ}}^{\min}$ distribution peaking at all three injected scan masses ($1000,\,1500,\,2000$~GeV) within $10\%$, together with a plotted deliverable. \\
\hline
\end{tabular}
\caption{Stagewise rubric for the leptoquark benchmark. Weights sum to unity. Half the credit is placed on the final physics deliverable, with the remainder distributed across the upstream machinery so that partial progress is resolved.}
\label{tab:rubric}
\end{table}

\begin{figure}[t]
\centering
\includegraphics[width=1.0\linewidth]{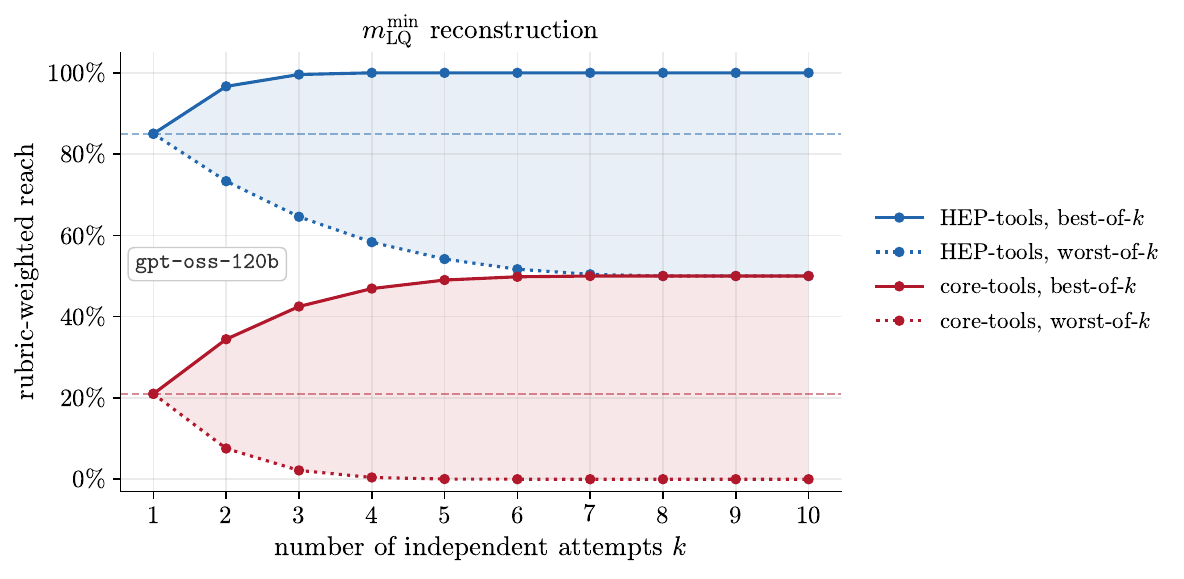}
\caption{Reach analogue of \cref{fig:passk} (GPT-OSS-120B, $N=10$). The figure shows rubric-weighted reach versus the number of attempts $k$ for the tool-assisted (blue) and core-only (red) variants. Solid curves show best-of-$k$ reach@$k$, dotted curves show worst-of-$k$ reach$^k$, and dashed lines show the single-shot mean. The worst-of-$k$ tool-assisted reach saturates near $0.5$, matching the best-of-$k$ core-only reach. The core-only ceiling coincides with the tool-assisted floor.}
\label{fig:reachk}
\end{figure}

Two quantities are derived from the per-stage outcomes. The \emph{reach} of trial $j$ is the rubric-weighted fraction of the pipeline it completes,
\begin{equation}
R_j \;=\; \frac{1}{W}\sum_{i=1}^{M} w_i \prod_{\ell \le i} S_{j,\ell}, \qquad W = \sum_{i=1}^{M} w_i,
\label{eq:reach}
\end{equation}
where $M$ is the number of rubric stages, $S_{j,i}\in\{0,1\}$ is the outcome of stage $i$ on trial $j$, and the cumulative product enforces the absorbing convention. A stage contributes only if it and every prior stage passed. A run \emph{passes} when it clears every stage. Equivalently, it passes when it produces a correct mass reconstruction, since that stage is downstream of all others. The pass@$k$ and pass$^k$ metrics therefore follow from the per-trial pass/fail outcomes as in \cref{sec:benchmarking}. The reported reach is the trial mean $\overline{R}$. Its best- and worst-of-$k$ aggregates $\text{reach@}k=\mathbb{E}[\max_{j\le k}R_j]$ and $\text{reach}^k=\mathbb{E}[\min_{j\le k}R_j]$ are the reach analogues of pass@$k$ and pass$^k$ (\cref{fig:reachk}). The content checks are deliberately permissive about format and location so that an agent is never penalized for saving a valid result under an unexpected name. Correctness of the final observable is enforced by the peak-position criterion of the last stage.

\begin{figure}[h!]
\centering
\includegraphics[width=0.72\linewidth]{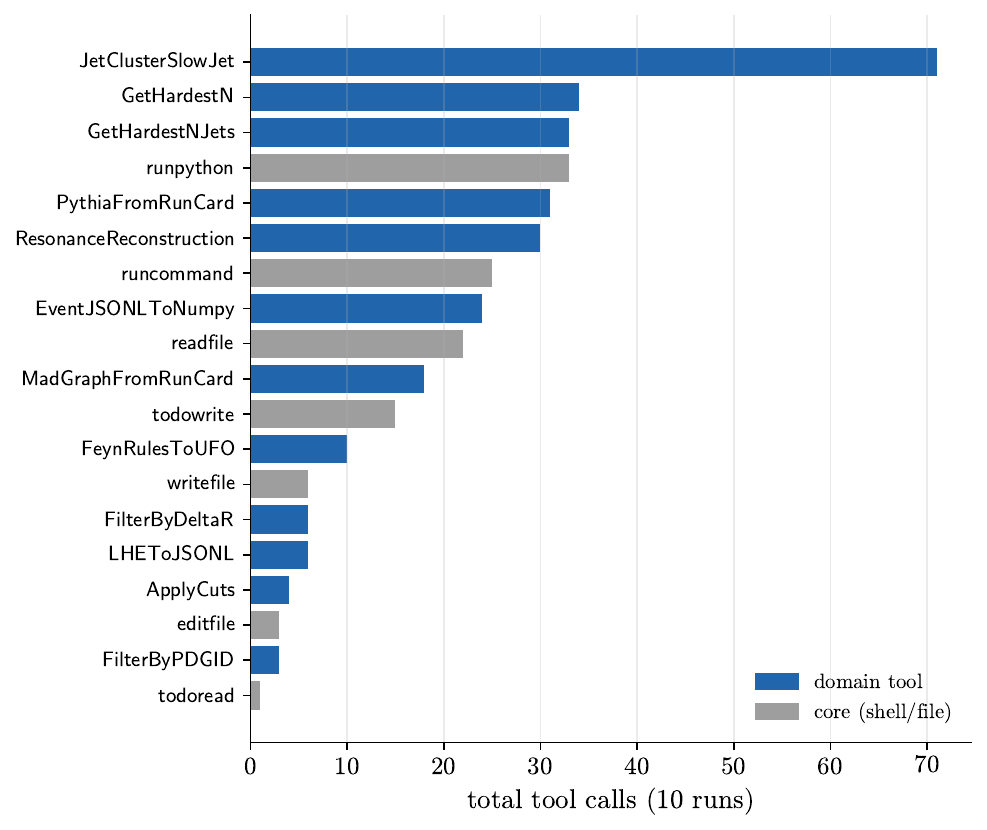}
\caption{Tool-usage histogram for the tool-assisted agent: total invocations of each tool summed over
the $N=10$ runs (baseline variant), with domain (HEPTAPOD) tools in blue and core shell/file
tools in grey.}
\label{fig:toolusage}
\end{figure}

\paragraph{Run-level characterization.}
\Cref{tab:runstats} breaks the two variants down run by run, and \cref{fig:toolusage} shows how often
the tool-assisted agent invokes each tool. Two points stand out. First, the serialization failure mode is
visible directly. Tool-assisted runs occasionally trigger a format-crash retry, but these retries are
never fatal because the structured tool arguments are small. Core-only runs, which must emit whole
scripts as JSON arguments, crash far more often and frequently fatally. Second, the agent spends the
majority of its calls on the domain tools, driving the full \textsc{FeynRules}$\,\to\,$\textsc{MadGraph}$\,\to\,$\textsc{Pythia}
$\,\to\,$clustering$\,\to\,$reconstruction chain rather than falling back on generic shell commands.

\begin{table}[h!]
\centering
\small
\begin{tabular}{c cccc cccc}
\toprule
 & \multicolumn{4}{c}{Tool-assisted} & \multicolumn{4}{c}{Core-only} \\
\cmidrule(lr){2-5}\cmidrule(lr){6-9}
run & reach & calls & tok & crash & reach & calls & tok & crash \\
\midrule
1 & $m_{\mathrm{LQ}}$ & 64 & 33\,k & 0 & reco & 31 & 7\,k & 2 \\
2 & reco & 29 & 11\,k & 0 & UFO & 21 & 8\,k & 2 \\
3 & $m_{\mathrm{LQ}}$ & 39 & 14\,k & 0 & n/a & 42 & 18\,k & 2 \\
4 & $m_{\mathrm{LQ}}$ & 27 & 11\,k & 0 & reco & 27 & 10\,k & 0 \\
5 & $m_{\mathrm{LQ}}$ & 34 & 14\,k & 0 & n/a & 15 & 4\,k & 2 \\
6 & reco & 36 & 12\,k & 0 & n/a & 9 & 4\,k & 2 \\
7 & reco & 29 & 12\,k & 0 & n/a & 43 & 16\,k & 2 \\
8 & $m_{\mathrm{LQ}}$ & 29 & 10\,k & 2 & n/a & 88 & 40\,k & 2 \\
9 & $m_{\mathrm{LQ}}$ & 53 & 26\,k & 1 & reco & 25 & 9\,k & 2 \\
10 & $m_{\mathrm{LQ}}$ & 35 & 15\,k & 0 & reco & 68 & 28\,k & 2 \\
\midrule
mean/tot & 7/10 & 38 & 16\,k & 2/10 & 0/10 & 37 & 14\,k & 9/10 \\
\bottomrule
\end{tabular}
\caption{Per-run breakdown for the two variants, seed-paired ($N=10$ each, GPT-OSS-120B).
The tool-assisted rows use the spec-matched \texttt{baseline} variant, and the core-only rows use the conservative \texttt{recipes} variant. \emph{reach} is the furthest pipeline
stage completed ($m_{\mathrm{LQ}}$ = correct resonance reconstruction, \emph{reco} = mass array only,
\emph{UFO}/\emph{parton}/\ldots\ = stopped earlier, n/a = no stage). \emph{calls} is the number of
tool calls. \emph{tok} is output tokens (thousands). \emph{crash} is the number of JSON format-crash
retries used (0--2). The \emph{mean/tot} row gives success rate, mean calls, mean tokens, and the number
of runs with at least one crash. Call and token counts characterize each variant and are not a like-for-like
efficiency comparison, since the core-only runs mostly terminate early.}
\label{tab:runstats}
\end{table}

\section{Tools}
\label{app:tools}

This appendix summarizes the domain-specific tools made available to the agent.
Tools are grouped by their role in the workflow and provide a uniform, schema-validated interface.
All tools inherit from the \texttt{BaseTool} abstraction, which defines the interface between tools and the LLM orchestrator by converting type-annotated Python classes into LLM-compatible function schemas, separating runtime parameters (\texttt{RuntimeField}) from configuration state (\texttt{StateField}), and providing automatic validation and structured error handling (see \cref{sec:architecture} and \cite{orchestral-ai} for details).
Each tool entry below specifies its \emph{Purpose}, \emph{Inputs}, \emph{Behavior}, and relevant \emph{Notes}.
Full implementations and API documentation are available in the public repository.

\begin{table*}[p!]
\centering
\begin{sideways}
\begin{tabular}{llp{10cm}}
\hline
\textbf{Category} & \textbf{Tool} & \textbf{Purpose} \\
\hline
\multirow{3}{*}{Event Generation}
  & \texttt{FeynRulesToUFOTool} & Generate a validated UFO model directory from a \textsc{FeynRules} \texttt{.fr} file for use in matrix-element generators. \\
  & \texttt{MadGraphFromRunCardTool} & Execute \textsc{MadGraph5\_aMC@NLO} with runtime substitutions and automatic identification of parameter scans. \\
  & \texttt{PythiaFromRunCardTool} & Run \textsc{Pythia}~8 event generation or LHE showering/hadronization with JSONL event output. \\
  & \texttt{SherpaFromRunCardTool} & Run \textsc{Sherpa}~3 event generation (including UFO conversion) with JSONL event output. \\
\hline
\multirow{3}{*}{Data Conversion}
  & \texttt{LHEToJSONLTool} & Convert LHE events into the unified \texttt{evtjsonl-1.0} schema used throughout the workflow. \\
  & \texttt{EventJSONLToNumpyTool} & Convert particle-level JSONL into padded NumPy arrays suitable for ML and vectorized analysis. \\
  & \texttt{JetsJSONLToNumpyTool} & Convert jet-level JSONL collections into fixed-shape NumPy arrays with configurable feature sets. \\
\hline
Jet Clustering
  & \texttt{JetClusterSlowJetTool} & Cluster jets using anti-$k_T$/CA/$k_T$ algorithms via \textsc{Pythia}'s \texttt{SlowJet} interface. \\
\hline
\multirow{7}{*}{Kinematic Analysis}
  & \texttt{CalculateInvariantMassTool} & Compute invariant masses for user-specified combinations of particles. \\
  & \texttt{CalculateTransverseMomentumTool} & Extract transverse-momentum distributions from particle or jet collections. \\
  & \texttt{CalculateDeltaRTool} & Compute $\Delta R$ separations between objects in one or two collections. \\
  & \texttt{ApplyCutsTool} & Apply kinematic selections with user-defined thresholds. \\
  & \texttt{SortByPtTool} & Order objects by transverse momentum within each event. \\
  & \texttt{MergeObjectCollectionsTool} & Merge multiple JSONL object collections into a single event-aligned file. \\
  & \texttt{FilterByDeltaRTool} & Remove objects based on $\Delta R$ proximity to another collection. \\
\hline
\multirow{3}{*}{Event Selection}
  & \texttt{GetHardestNTool} & Select the $N$ highest-$p_T$ particles per event with optional species filtering. \\
  & \texttt{GetHardestNJetsTool} & Select the $N$ highest-$p_T$ jets per event from jet JSONL files. \\
  & \texttt{FilterByPDGIDTool} & Retain only particles with specified PDG IDs and report filtering efficiencies. \\
\hline
Advanced Analysis
  & \texttt{ResonanceReconstructionTool} & Perform physics-aware two-body mass reconstruction from multiple object collections. \\
\hline
\end{tabular}
\end{sideways}
\caption{Summary of the HEP orchestration tool suite exposed to the agent.}
\label{tab:tools}
\end{table*}


\subsection{Event Generation}

\paragraph{\texttt{FeynRulesToUFOTool}.}
\textbf{Purpose.} Produce a complete and validated UFO model directory from a \textsc{FeynRules} \texttt{.fr} file, enabling subsequent use in \textsc{MadGraph}.
\textbf{Inputs.} Model file path, UFO output directory, optional \texttt{feynrules\_path}, \texttt{wolframscript} path, timeout.
\textbf{Behavior.} Executes an isolated \textsc{Mathematica} session to load the model and generate the UFO; verifies the resulting directory structure (particles, parameters, vertices, couplings).
\textbf{Notes.} Errors originating from model syntax or missing executables are surfaced with structured metadata to support agent-level recovery.

\paragraph{\texttt{MadGraphFromRunCardTool}.}
\textbf{Purpose.} Execute \textsc{MadGraph5\_aMC@NLO} using a template run card with runtime modifications, and present parameter scans in a uniform structured format.  
\textbf{Inputs.} Template \texttt{.mg5} file and optional overrides: UFO path, output name, event count, seed.  
\textbf{Behavior.} Rewrites the run card with substitutions, runs MG5, detects multi-run scans from MG5 output directories, and returns structured metadata per run (parameter values, LHE paths, cross sections, uncertainties).  
\textbf{Notes.} Single-run and multi-run cards are exposed to the agent through the same metadata schema.

\paragraph{\texttt{PythiaFromRunCardTool}.}
\textbf{Purpose.} Provide \textsc{Pythia}~8 event generation, including hard-process simulation, parton showering, and hadronization, with optional LHE input and standardized JSONL event output. Supports both fully standalone generation and LHE-based showering/hadronizing modes.
\textbf{Inputs.} Command card, number of events, seed, flags for final-state filtering, optional \texttt{shower\_lhe} flag and \texttt{lhe\_path}.
\textbf{Behavior.} Runs Pythia with the supplied configuration; outputs each event in the \texttt{evtjsonl-1.0} schema and records cross-section information from Pythia's statistics block. In LHE showering mode (\texttt{shower\_lhe=True}), applies parton showering and hadronization to pre-generated matrix-element events.
\textbf{Notes.} When showering LHE files, the tool automatically injects the appropriate \texttt{Beams:LHEF} directive with the absolute path to the LHE file.

\paragraph{\texttt{SherpaFromRunCardTool}.}
\textbf{Purpose.} Provide \textsc{Sherpa}~3 event generation, including hard-process simulation, parton showering, and hadronization, with direct calculation of all matrix elements based and standardized JSONL event output. Supports standalone generation with an interface to the UFO output from the \texttt{FeynRulesToUFOTool}.
\textbf{Inputs.} Run card, number of events, seed, flags for final-state filtering, optional UFO path.
\textbf{Behavior.} Runs Sherpa with the supplied configuration; outputs each event in the \texttt{evtjsonl-1.0} schema and records cross-section information from Sherpa's event weight information.
\textbf{Notes.} When in UFO mode, converts UFO model to Sherpa syntax and compiles associated dynamic libraries as needed.


\subsection{Data Conversion}

\paragraph{\texttt{LHEToJSONLTool}.}
\textbf{Purpose.} Convert LHE events into the unified JSONL event format used across the orchestration workflow.  
\textbf{Inputs.} LHE file path, output JSONL file path.  
\textbf{Behavior.} Parses all events and emits line-delimited JSON objects following the \texttt{evtjsonl-1.0} schema.  
\textbf{Notes.} Serves as a standardized bridge between matrix-element generators and downstream tools.

\paragraph{\texttt{EventJSONLToNumpyTool}.}
\textbf{Purpose.} Provide a fixed-shape NumPy representation of particle-level events for batch or ML processing.
\textbf{Inputs.} JSONL file, output directory.
\textbf{Behavior.} Identifies the maximum particle multiplicity and constructs an array of shape $(N_{\mathrm{events}}, N_{\mathrm{max}}, 5)$ with columns $[p_x,p_y,p_z,E,\texttt{id}]$ where $\texttt{id}$ is the PDG identifier.
\textbf{Notes.} Ensures deterministic shapes compatible with JAX, PyTorch, and NumPy.

\paragraph{\texttt{JetsJSONLToNumpyTool}.}
\textbf{Purpose.} Convert jet collections to fixed-shape NumPy arrays with configurable kinematic features.  
\textbf{Inputs.} Jet JSONL file, output directory, feature-extraction mode.  
\textbf{Behavior.} Extracts per-jet kinematics and optional auxiliary information; pads events to consistent shapes.  
\textbf{Notes.} Supports ML workflows requiring homogeneous jet representations.


\subsection{Jet Clustering}

\paragraph{\texttt{JetClusterSlowJetTool}.}
\textbf{Purpose.} Perform jet clustering using anti-$k_T$, Cambridge/Aachen, or $k_T$ algorithms through \textsc{Pythia}'s \texttt{SlowJet}.
\textbf{Inputs.} Particle JSONL file, algorithm choice, radius $R$, $p_T$ threshold, $\eta$ range, \texttt{cluster\_all} flag, optional output path.
\textbf{Behavior.} Constructs Pythia event containers from JSONL inputs, applies the chosen algorithm, and outputs jet collections in JSONL format with full kinematics and constituent indices. When \texttt{cluster\_all=True}, processes entire dataset and streams results to the specified output file.
\textbf{Notes.} Supports both single-event and batch-processing modes for large datasets.


\subsection{Kinematic Analysis}

\paragraph{\texttt{CalculateInvariantMassTool}.}
\textbf{Purpose.} Compute invariant masses for user-specified particle combinations within each event.  
\textbf{Inputs.} Particle JSONL file, index selections.  
\textbf{Behavior.} Computes $m=\sqrt{E^2 - \vec p^{\,2}}$ for each combination and writes results to JSON.  
\textbf{Notes.} Integrates with selection tools for higher-level analyses.

\paragraph{\texttt{CalculateTransverseMomentumTool}.}
\textbf{Purpose.} Extract particle or jet transverse-momentum distributions.  
\textbf{Inputs.} JSONL file.  
\textbf{Behavior.} Computes $p_T=\sqrt{p_x^2+p_y^2}$ for all objects and outputs per-event lists.  

\paragraph{\texttt{CalculateDeltaRTool}.}
\textbf{Purpose.} Compute angular separation between objects in one or two collections.  
\textbf{Inputs.} One or two JSONL collections; optional matching strategies.  
\textbf{Behavior.} Returns $\Delta R$ for all relevant pairs.  

\paragraph{\texttt{ApplyCutsTool}.}
\textbf{Purpose.} Apply user-defined kinematic cuts to particle or jet collections.  
\textbf{Inputs.} JSONL file, cut definitions.  
\textbf{Behavior.} Evaluates cuts per event and outputs the filtered JSONL file.  

\paragraph{\texttt{SortByPtTool}.}
\textbf{Purpose.} Sort objects by transverse momentum within each event.  
\textbf{Inputs.} JSONL file.  
\textbf{Behavior.} Computes $p_T$ and sorts objects in descending order.  

\paragraph{\texttt{MergeObjectCollectionsTool}.}
\textbf{Purpose.} Combine multiple object collections into a single event-aligned JSONL file.  
\textbf{Inputs.} List of JSONL files.  
\textbf{Behavior.} Merges per-event objects while preserving event IDs.  

\paragraph{\texttt{FilterByDeltaRTool}.}
\textbf{Purpose.} Remove objects based on $\Delta R$ proximity to objects in other collections, enabling overlap removal and isolation criteria.
\textbf{Inputs.} List of JSONL files containing pre-selected objects, $\Delta R$ threshold, filter mode (\texttt{remove\_first}, \texttt{remove\_second}, \texttt{remove\_both}, \texttt{keep\_only\_separated}), optional output paths.
\textbf{Behavior.} Applies $\Delta R$ constraints only between objects from different input arrays; objects within the same array are never constrained by this criterion. Permanently removes objects that fail the separation requirements and outputs filtered collections.
\textbf{Notes.} Common use cases include jet-lepton overlap removal and lepton isolation. Distinct from \texttt{ResonanceReconstructionTool}: this tool permanently removes objects and is not suitable for analyses requiring object pairing such as resonance mass reconstruction.  


\subsection{Event Selection}

\paragraph{\texttt{GetHardestNTool}.}
\textbf{Purpose.} Select the $N$ highest-$p_T$ particles per event, optionally restricted by PDG ID.
\textbf{Inputs.} Particle JSONL file, $N$, optional PDG list.
\textbf{Behavior.} Computes $p_T$, ranks objects, and retains the top $N$.
\textbf{Notes.} This is a generic leading-object selection, not specific to any analysis; hardest-$N$ selection is ubiquitous across BSM searches, SM measurements, and detector studies.

\paragraph{\texttt{GetHardestNJetsTool}.}
\textbf{Purpose.} Select the $N$ highest-$p_T$ jets per event from jet JSONL files.
\textbf{Inputs.} Jet JSONL file (with \texttt{data.jets} schema from \texttt{JetClusterSlowJetTool}), $N$.
\textbf{Behavior.} Orders jets by $p_T$ and retains the leading $N$.
\textbf{Notes.} Implements the same generic leading-object selection as \texttt{GetHardestNTool}, specialized only to the jet data type (\texttt{data.jets}) rather than to any particular analysis. For particle-level objects, use \texttt{GetHardestNTool} instead.

\paragraph{\texttt{FilterByPDGIDTool}.}
\textbf{Purpose.} Filter particle collections to retain only specified PDG IDs.  
\textbf{Inputs.} JSONL file, PDG list.  
\textbf{Behavior.} Applies per-event filtering and reports efficiencies.  


\subsection{Advanced Analysis}

\paragraph{\texttt{ResonanceReconstructionTool}.}
\textbf{Purpose.} Perform $n$-body resonance reconstruction from multiple object collections using physics-aware pairing templates.
\textbf{Inputs.} Lists of JSONL files (e.g., leptons and jets), optional minimum $\Delta R$ separation (\texttt{min\_delta\_r}), mass-pairing template (\texttt{two\_body\_symmetric} or \texttt{n\_body\_all\_pairs}).
\textbf{Behavior.} Evaluates all allowed disjoint pairings consistent with the template, selects the pairing minimizing $|m_1 - m_2|$ (for symmetric template), and outputs per-event mass observables in \texttt{.npy} format with summary statistics and histograms.
\textbf{Notes.} When \texttt{min\_delta\_r} is specified, $\Delta R$ constraints apply \emph{only across input collections} (e.g., lepton-jet separation), not within collections (e.g., lepton-lepton or jet-jet pairs are unconstrained). Pairings that combine only same-collection objects are automatically rejected when multiple collections are provided with active $\Delta R$ constraints, ensuring physically meaningful resonance reconstruction.

\end{appendix}

\newpage
\bibliography{ref}{}
\bibliographystyle{JHEP}

\end{document}

%% file: orchestral.tex
\usepackage[most]{tcolorbox}
\usepackage{minted}
\usepackage{fancyvrb}  
\usepackage{xcolor}
\usepackage{xspace}
\usepackage{setspace}
\usepackage{enumitem}  
\usepackage{etoolbox}  
\usepackage{varwidth}  
\definecolor{darkblue}{RGB}{0,0,139}

\newcommand{\orchestralai}{\textsc{Orchestral AI}\xspace}

\newcommand{\orchestralnestindent}{1.5em}  

\newcommand{\orchestralfontsize}{\small}  

\newcommand{\orchestralcodestyle}{vs}  

\newcommand{\orchestralitemsep}{0pt}           
\newcommand{\orchestralparsep}{0pt}            
\newcommand{\orchestraltopsep}{0pt}            
\newcommand{\orchestralpartopsep}{0pt}         

\definecolor{OrcUserColor}{RGB}{210, 80, 80}         
\definecolor{OrcAgentColor}{RGB}{120, 120, 255}     
\definecolor{OrcSystemColor}{RGB}{180, 30, 180}      
\definecolor{OrcToolColor}{RGB}{200, 100, 0}        

\definecolor{OrcPendingColor}{RGB}{128, 128, 128}   
\definecolor{OrcFailedColor}{RGB}{220, 20, 60}      

\definecolor{OrcCodeBg}{RGB}{36, 41, 51}            

\newtcolorbox{orchestralusermessage}{%
  enhanced,
  breakable,
  colback=white,
  colframe=OrcUserColor,
  boxrule=0.6pt,
  arc=1.5mm,
  width=\linewidth,
  top=8pt,
  bottom=8pt,
  left=10pt,
  right=10pt,
  fontupper=\orchestralfontsize,
  attach boxed title to top left={
    yshift=-6pt,
    xshift=3.5mm
  },
  boxed title style={
    colback=white,
    colframe=white,
    boxrule=0pt,
    left=0pt,
    right=0pt,
    top=3pt,
    bottom=0pt,
    sharp corners,
  },
  coltitle=OrcUserColor,
  fonttitle=\sffamily,
  title={\raisebox{0pt}[0pt][0pt]{User}},
}

\newtcolorbox{orchestralagentmessage}{%
  enhanced,
  breakable,
  colback=white,
  colframe=OrcAgentColor,
  boxrule=0.6pt,
  arc=1.5mm,
  width=\linewidth,
  top=8pt,
  bottom=8pt,
  left=10pt,
  right=10pt,
  fontupper=\orchestralfontsize,
  attach boxed title to top left={
    yshift=-6pt,
    xshift=3.5mm
  },
  boxed title style={
    colback=white,
    colframe=white,
    boxrule=0pt,
    left=0pt,
    right=0pt,
    top=3pt,
    bottom=0pt,
    sharp corners,
  },
  coltitle=OrcAgentColor,
  fonttitle=\sffamily,
  title={\raisebox{0pt}[0pt][0pt]{Agent}},
}

\newtcolorbox{orchestralsystemmessage}{%
  enhanced,
  breakable,
  colback=white,
  colframe=OrcSystemColor,
  boxrule=0.6pt,
  arc=1.5mm,
  width=\linewidth,
  top=8pt,
  bottom=8pt,
  left=10pt,
  right=10pt,
  fontupper=\orchestralfontsize,
  attach boxed title to top left={
    yshift=-6pt,
    xshift=3.5mm
  },
  boxed title style={
    colback=white,
    colframe=white,
    boxrule=0pt,
    left=0pt,
    right=0pt,
    top=3pt,
    bottom=0pt,
    sharp corners,
  },
  coltitle=OrcSystemColor,
  fonttitle=\sffamily,
  title={\raisebox{0pt}[0pt][0pt]{System}},
}

\newtcolorbox{orchestraltoolmessage}[1]{%
  enhanced,
  breakable,
  colback=white,
  colframe=OrcToolColor,
  boxrule=0.5pt,
  arc=1mm,
  width=\linewidth,
  left=\orchestralnestindent,
  right=1em,
  top=6pt,
  bottom=6pt,
  fontupper=\orchestralfontsize,
  attach boxed title to top left={
    yshift=-5pt,
    xshift=3.5mm
  },
  boxed title style={
    colback=white,
    colframe=white,
    boxrule=0pt,
    left=0pt,
    right=0pt,
    top=3pt,
    bottom=0pt,
    sharp corners,
  },
  coltitle=OrcToolColor,
  fonttitle=\sffamily\orchestralfontsize,
  title={\raisebox{0pt}[0pt][0pt]{#1}},
}

\newtcolorbox{orchestraltoolmessageml}[1]{%
  enhanced,
  breakable,
  colback=white,
  colframe=OrcToolColor,
  boxrule=0.5pt,
  arc=1mm,
  width=\linewidth,
  left=\orchestralnestindent,
  right=1em,
  top=34pt,
  bottom=6pt,
  fontupper=\orchestralfontsize,
  attach boxed title to top left={
    yshift=-35pt,
    xshift=3.5mm
  },
  boxed title style={
    colback=white,
    colframe=white,
    boxrule=0pt,
    left=0pt,
    right=0pt,
    top=3pt,
    bottom=0pt,
    sharp corners,
  },
  coltitle=OrcToolColor,
  fonttitle=\sffamily\orchestralfontsize,
  title={\begin{varwidth}{\dimexpr\linewidth-\orchestralnestindent-1em-3.5mm\relax}#1\end{varwidth}},
}

\newminted{python}{
  style=\orchestralcodestyle,
  fontsize=\small,
  breaklines,
  autogobble
}

\newminted{bash}{
  style=\orchestralcodestyle,
  fontsize=\small,
  breaklines,
  autogobble
}

\setlist[enumerate]{
  itemsep=\orchestralitemsep,
  parsep=\orchestralparsep,
  topsep=\orchestraltopsep,
  partopsep=\orchestralpartopsep
}

\setlist[itemize]{
  itemsep=\orchestralitemsep,
  parsep=\orchestralparsep,
  topsep=\orchestraltopsep,
  partopsep=\orchestralpartopsep
}

\setminted{
  style=vs,
  fontsize=\orchestralfontsize,
  frame=lines,
  bgcolor=gray!5,
  xleftmargin=0pt,
  xrightmargin=0pt,
  tabsize=4,
  showtabs=false
}

\AtBeginEnvironment{minted}{\vspace{0.3\baselineskip}}
\AtEndEnvironment{minted}{\vspace{0.3\baselineskip}}

%% file: ref.bib
@article{CaloChallenge2024,
    author = "Krause, Claudius and others",
    collaboration = "CaloChallenge",
    title = "{CaloChallenge 2022: A Community Challenge for Fast Calorimeter Simulation}",
    eprint = "2410.21611",
    archivePrefix = "arXiv",
    primaryClass = "cs.LG",
    year = "2024"
}

@article{zhu2025heptapodlanguagemodelingvisual,
      title={Heptapod: Language Modeling on Visual Signals}, 
      author={Yongxin Zhu and Jiawei Chen and Yuanzhe Chen and Zhuo Chen and Dongya Jia and Jian Cong and Xiaobin Zhuang and Yuping Wang and Yuxuan Wang},
      year={2025},
      eprint={2510.06673},
      archivePrefix={arXiv},
      primaryClass={cs.CV},
      url={https://arxiv.org/abs/2510.06673}, 
}

@article{orchestral-ai,
      title={Orchestral AI: A Framework for Agent Orchestration}, 
      author={Alexander Roman and Jacob Roman},
      year={2026},
      eprint={2601.02577},
      archivePrefix={arXiv},
      primaryClass={cs.AI},
      url={https://arxiv.org/abs/2601.02577}, 
}

@article{panek2026asteragenticscience,
      title={ASTER -- Agentic Science Toolkit for Exoplanet Research}, 
      author={Emilie Panek and Alexander Roman and Gaurav Shukla and Leonardo Pagliaro and Katia Matcheva and Konstantin Matchev},
      year={2026},
      eprint={2603.26953},
      archivePrefix={arXiv},
      primaryClass={astro-ph.EP},
      url={https://arxiv.org/abs/2603.26953}, 
}

@article{openai2024gpt4technicalreport,
      title={GPT-4 Technical Report}, 
      author={OpenAI and Josh Achiam and Steven Adler and Sandhini Agarwal and Lama Ahmad and Ilge Akkaya and Florencia Leoni Aleman and Diogo Almeida and Janko Altenschmidt and Sam Altman and Shyamal Anadkat and Red Avila and Igor Babuschkin and Suchir Balaji and Valerie Balcom and Paul Baltescu and Haiming Bao and Mohammad Bavarian and Jeff Belgum and Irwan Bello and Jake Berdine and Gabriel Bernadett-Shapiro and Christopher Berner and Lenny Bogdonoff and Oleg Boiko and Madelaine Boyd and Anna-Luisa Brakman and Greg Brockman and Tim Brooks and Miles Brundage and Kevin Button and Trevor Cai and Rosie Campbell and Andrew Cann and Brittany Carey and Chelsea Carlson and Rory Carmichael and Brooke Chan and Che Chang and Fotis Chantzis and Derek Chen and Sully Chen and Ruby Chen and Jason Chen and Mark Chen and Ben Chess and Chester Cho and Casey Chu and Hyung Won Chung and Dave Cummings and Jeremiah Currier and Yunxing Dai and Cory Decareaux and Thomas Degry and Noah Deutsch and Damien Deville and Arka Dhar and David Dohan and Steve Dowling and Sheila Dunning and Adrien Ecoffet and Atty Eleti and Tyna Eloundou and David Farhi and Liam Fedus and Niko Felix and Simón Posada Fishman and Juston Forte and Isabella Fulford and Leo Gao and Elie Georges and Christian Gibson and Vik Goel and Tarun Gogineni and Gabriel Goh and Rapha Gontijo-Lopes and Jonathan Gordon and Morgan Grafstein and Scott Gray and Ryan Greene and Joshua Gross and Shixiang Shane Gu and Yufei Guo and Chris Hallacy and Jesse Han and Jeff Harris and Yuchen He and Mike Heaton and Johannes Heidecke and Chris Hesse and Alan Hickey and Wade Hickey and Peter Hoeschele and Brandon Houghton and Kenny Hsu and Shengli Hu and Xin Hu and Joost Huizinga and Shantanu Jain and Shawn Jain and Joanne Jang and Angela Jiang and Roger Jiang and Haozhun Jin and Denny Jin and Shino Jomoto and Billie Jonn and Heewoo Jun and Tomer Kaftan and Łukasz Kaiser and Ali Kamali and Ingmar Kanitscheider and Nitish Shirish Keskar and Tabarak Khan and Logan Kilpatrick and Jong Wook Kim and Christina Kim and Yongjik Kim and Jan Hendrik Kirchner and Jamie Kiros and Matt Knight and Daniel Kokotajlo and Łukasz Kondraciuk and Andrew Kondrich and Aris Konstantinidis and Kyle Kosic and Gretchen Krueger and Vishal Kuo and Michael Lampe and Ikai Lan and Teddy Lee and Jan Leike and Jade Leung and Daniel Levy and Chak Ming Li and Rachel Lim and Molly Lin and Stephanie Lin and Mateusz Litwin and Theresa Lopez and Ryan Lowe and Patricia Lue and Anna Makanju and Kim Malfacini and Sam Manning and Todor Markov and Yaniv Markovski and Bianca Martin and Katie Mayer and Andrew Mayne and Bob McGrew and Scott Mayer McKinney and Christine McLeavey and Paul McMillan and Jake McNeil and David Medina and Aalok Mehta and Jacob Menick and Luke Metz and Andrey Mishchenko and Pamela Mishkin and Vinnie Monaco and Evan Morikawa and Daniel Mossing and Tong Mu and Mira Murati and Oleg Murk and David Mély and Ashvin Nair and Reiichiro Nakano and Rajeev Nayak and Arvind Neelakantan and Richard Ngo and Hyeonwoo Noh and Long Ouyang and Cullen O'Keefe and Jakub Pachocki and Alex Paino and Joe Palermo and Ashley Pantuliano and Giambattista Parascandolo and Joel Parish and Emy Parparita and Alex Passos and Mikhail Pavlov and Andrew Peng and Adam Perelman and Filipe de Avila Belbute Peres and Michael Petrov and Henrique Ponde de Oliveira Pinto and Michael and Pokorny and Michelle Pokrass and Vitchyr H. Pong and Tolly Powell and Alethea Power and Boris Power and Elizabeth Proehl and Raul Puri and Alec Radford and Jack Rae and Aditya Ramesh and Cameron Raymond and Francis Real and Kendra Rimbach and Carl Ross and Bob Rotsted and Henri Roussez and Nick Ryder and Mario Saltarelli and Ted Sanders and Shibani Santurkar and Girish Sastry and Heather Schmidt and David Schnurr and John Schulman and Daniel Selsam and Kyla Sheppard and Toki Sherbakov and Jessica Shieh and Sarah Shoker and Pranav Shyam and Szymon Sidor and Eric Sigler and Maddie Simens and Jordan Sitkin and Katarina Slama and Ian Sohl and Benjamin Sokolowsky and Yang Song and Natalie Staudacher and Felipe Petroski Such and Natalie Summers and Ilya Sutskever and Jie Tang and Nikolas Tezak and Madeleine B. Thompson and Phil Tillet and Amin Tootoonchian and Elizabeth Tseng and Preston Tuggle and Nick Turley and Jerry Tworek and Juan Felipe Cerón Uribe and Andrea Vallone and Arun Vijayvergiya and Chelsea Voss and Carroll Wainwright and Justin Jay Wang and Alvin Wang and Ben Wang and Jonathan Ward and Jason Wei and CJ Weinmann and Akila Welihinda and Peter Welinder and Jiayi Weng and Lilian Weng and Matt Wiethoff and Dave Willner and Clemens Winter and Samuel Wolrich and Hannah Wong and Lauren Workman and Sherwin Wu and Jeff Wu and Michael Wu and Kai Xiao and Tao Xu and Sarah Yoo and Kevin Yu and Qiming Yuan and Wojciech Zaremba and Rowan Zellers and Chong Zhang and Marvin Zhang and Shengjia Zhao and Tianhao Zheng and Juntang Zhuang and William Zhuk and Barret Zoph},
      year={2024},
      eprint={2303.08774},
      archivePrefix={arXiv},
      primaryClass={cs.CL},
      url={https://arxiv.org/abs/2303.08774}, 
}

@article{bubeck2023sparks,
      title={Sparks of Artificial General Intelligence: Early experiments with GPT-4}, 
      author={Sébastien Bubeck and Varun Chandrasekaran and Ronen Eldan and Johannes Gehrke and Eric Horvitz and Ece Kamar and Peter Lee and Yin Tat Lee and Yuanzhi Li and Scott Lundberg and Harsha Nori and Hamid Palangi and Marco Tulio Ribeiro and Yi Zhang},
      year={2023},
      eprint={2303.12712},
      archivePrefix={arXiv},
      primaryClass={cs.CL},
      url={https://arxiv.org/abs/2303.12712}, 
}

@article{zhao2023survey,
      title={A Survey of Large Language Models}, 
      author={Wayne Xin Zhao and Kun Zhou and Junyi Li and Tianyi Tang and Xiaolei Wang and Yupeng Hou and Yingqian Min and Beichen Zhang and Junjie Zhang and Zican Dong and Yifan Du and Chen Yang and Yushuo Chen and Zhipeng Chen and Jinhao Jiang and Ruiyang Ren and Yifan Li and Xinyu Tang and Zikang Liu and Peiyu Liu and Jian-Yun Nie and Ji-Rong Wen},
      year={2025},
      eprint={2303.18223},
      archivePrefix={arXiv},
      primaryClass={cs.CL},
      url={https://arxiv.org/abs/2303.18223}, 
}

@article{yin2023multimodal,
   title={A survey on multimodal large language models},
   volume={11},
   ISSN={2053-714X},
   url={http://dx.doi.org/10.1093/nsr/nwae403},
   DOI={10.1093/nsr/nwae403},
   number={12},
   journal={National Science Review},
   publisher={Oxford University Press (OUP)},
   author={Yin, Shukang and Fu, Chaoyou and Zhao, Sirui and Li, Ke and Sun, Xing and Xu, Tong and Chen, Enhong},
   year={2024},
   eprint={2306.13549},
   archivePrefix={arXiv},
   primaryClass={cs.CL},
   month=nov }

@article{minaee2025largelanguagemodelssurvey,
      title={Large Language Models: A Survey}, 
      author={Shervin Minaee and Tomas Mikolov and Narjes Nikzad and Meysam Chenaghlu and Richard Socher and Xavier Amatriain and Jianfeng Gao},
      year={2025},
      eprint={2402.06196},
      archivePrefix={arXiv},
      primaryClass={cs.CL},
      url={https://arxiv.org/abs/2402.06196}, 
}

@misc{grattafiori2024llama3herdmodels,
      title={The Llama 3 Herd of Models}, 
      author={Aaron Grattafiori and Abhimanyu Dubey and Abhinav Jauhri and Abhinav Pandey and Abhishek Kadian and Ahmad Al-Dahle and Aiesha Letman and Akhil Mathur and Alan Schelten and Alex Vaughan and Amy Yang and Angela Fan and Anirudh Goyal and Anthony Hartshorn and Aobo Yang and Archi Mitra and Archie Sravankumar and Artem Korenev and Arthur Hinsvark and Arun Rao and Aston Zhang and Aurelien Rodriguez and Austen Gregerson and Ava Spataru and Baptiste Roziere and Bethany Biron and Binh Tang and Bobbie Chern and Charlotte Caucheteux and Chaya Nayak and Chloe Bi and Chris Marra and Chris McConnell and Christian Keller and Christophe Touret and Chunyang Wu and Corinne Wong and Cristian Canton Ferrer and Cyrus Nikolaidis and Damien Allonsius and Daniel Song and Danielle Pintz and Danny Livshits and Danny Wyatt and David Esiobu and Dhruv Choudhary and Dhruv Mahajan and Diego Garcia-Olano and Diego Perino and Dieuwke Hupkes and Egor Lakomkin and Ehab AlBadawy and Elina Lobanova and Emily Dinan and Eric Michael Smith and Filip Radenovic and Francisco Guzmán and Frank Zhang and Gabriel Synnaeve and Gabrielle Lee and Georgia Lewis Anderson and Govind Thattai and Graeme Nail and Gregoire Mialon and Guan Pang and Guillem Cucurell and Hailey Nguyen and Hannah Korevaar and Hu Xu and Hugo Touvron and Iliyan Zarov and Imanol Arrieta Ibarra and Isabel Kloumann and Ishan Misra and Ivan Evtimov and Jack Zhang and Jade Copet and Jaewon Lee and Jan Geffert and Jana Vranes and Jason Park and Jay Mahadeokar and Jeet Shah and Jelmer van der Linde and Jennifer Billock and Jenny Hong and Jenya Lee and Jeremy Fu and Jianfeng Chi and Jianyu Huang and Jiawen Liu and Jie Wang and Jiecao Yu and Joanna Bitton and Joe Spisak and Jongsoo Park and Joseph Rocca and Joshua Johnstun and Joshua Saxe and Junteng Jia and Kalyan Vasuden Alwala and Karthik Prasad and Kartikeya Upasani and Kate Plawiak and Ke Li and Kenneth Heafield and Kevin Stone and Khalid El-Arini and Krithika Iyer and Kshitiz Malik and Kuenley Chiu and Kunal Bhalla and Kushal Lakhotia and Lauren Rantala-Yeary and Laurens van der Maaten and Lawrence Chen and Liang Tan and Liz Jenkins and Louis Martin and Lovish Madaan and Lubo Malo and Lukas Blecher and Lukas Landzaat and Luke de Oliveira and Madeline Muzzi and Mahesh Pasupuleti and Mannat Singh and Manohar Paluri and Marcin Kardas and Maria Tsimpoukelli and Mathew Oldham and Mathieu Rita and Maya Pavlova and Melanie Kambadur and Mike Lewis and Min Si and Mitesh Kumar Singh and Mona Hassan and Naman Goyal and Narjes Torabi and Nikolay Bashlykov and Nikolay Bogoychev and Niladri Chatterji and Ning Zhang and Olivier Duchenne and Onur Çelebi and Patrick Alrassy and Pengchuan Zhang and Pengwei Li and Petar Vasic and Peter Weng and Prajjwal Bhargava and Pratik Dubal and Praveen Krishnan and Punit Singh Koura and Puxin Xu and Qing He and Qingxiao Dong and Ragavan Srinivasan and Raj Ganapathy and Ramon Calderer and Ricardo Silveira Cabral and Robert Stojnic and Roberta Raileanu and Rohan Maheswari and Rohit Girdhar and Rohit Patel and Romain Sauvestre and Ronnie Polidoro and Roshan Sumbaly and Ross Taylor and Ruan Silva and Rui Hou and Rui Wang and Saghar Hosseini and Sahana Chennabasappa and Sanjay Singh and Sean Bell and Seohyun Sonia Kim and Sergey Edunov and Shaoliang Nie and Sharan Narang and Sharath Raparthy and Sheng Shen and Shengye Wan and Shruti Bhosale and Shun Zhang and Simon Vandenhende and Soumya Batra and Spencer Whitman and Sten Sootla and Stephane Collot and Suchin Gururangan and Sydney Borodinsky and Tamar Herman and Tara Fowler and Tarek Sheasha and Thomas Georgiou and Thomas Scialom and Tobias Speckbacher and Todor Mihaylov and Tong Xiao and Ujjwal Karn and Vedanuj Goswami and Vibhor Gupta and Vignesh Ramanathan and Viktor Kerkez and Vincent Gonguet and Virginie Do and Vish Vogeti and Vítor Albiero and Vladan Petrovic and Weiwei Chu and Wenhan Xiong and Wenyin Fu and Whitney Meers and Xavier Martinet and Xiaodong Wang and Xiaofang Wang and Xiaoqing Ellen Tan and Xide Xia and Xinfeng Xie and Xuchao Jia and Xuewei Wang and Yaelle Goldschlag and Yashesh Gaur and Yasmine Babaei and Yi Wen and Yiwen Song and Yuchen Zhang and Yue Li and Yuning Mao and Zacharie Delpierre Coudert and Zheng Yan and Zhengxing Chen and Zoe Papakipos and Aaditya Singh and Aayushi Srivastava and Abha Jain and Adam Kelsey and Adam Shajnfeld and Adithya Gangidi and Adolfo Victoria and Ahuva Goldstand and Ajay Menon and Ajay Sharma and Alex Boesenberg and Alexei Baevski and Allie Feinstein and Amanda Kallet and Amit Sangani and Amos Teo and Anam Yunus and Andrei Lupu and Andres Alvarado and Andrew Caples and Andrew Gu and Andrew Ho and Andrew Poulton and Andrew Ryan and Ankit Ramchandani and Annie Dong and Annie Franco and Anuj Goyal and Aparajita Saraf and Arkabandhu Chowdhury and Ashley Gabriel and Ashwin Bharambe and Assaf Eisenman and Azadeh Yazdan and Beau James and Ben Maurer and Benjamin Leonhardi and Bernie Huang and Beth Loyd and Beto De Paola and Bhargavi Paranjape and Bing Liu and Bo Wu and Boyu Ni and Braden Hancock and Bram Wasti and Brandon Spence and Brani Stojkovic and Brian Gamido and Britt Montalvo and Carl Parker and Carly Burton and Catalina Mejia and Ce Liu and Changhan Wang and Changkyu Kim and Chao Zhou and Chester Hu and Ching-Hsiang Chu and Chris Cai and Chris Tindal and Christoph Feichtenhofer and Cynthia Gao and Damon Civin and Dana Beaty and Daniel Kreymer and Daniel Li and David Adkins and David Xu and Davide Testuggine and Delia David and Devi Parikh and Diana Liskovich and Didem Foss and Dingkang Wang and Duc Le and Dustin Holland and Edward Dowling and Eissa Jamil and Elaine Montgomery and Eleonora Presani and Emily Hahn and Emily Wood and Eric-Tuan Le and Erik Brinkman and Esteban Arcaute and Evan Dunbar and Evan Smothers and Fei Sun and Felix Kreuk and Feng Tian and Filippos Kokkinos and Firat Ozgenel and Francesco Caggioni and Frank Kanayet and Frank Seide and Gabriela Medina Florez and Gabriella Schwarz and Gada Badeer and Georgia Swee and Gil Halpern and Grant Herman and Grigory Sizov and Guangyi and Zhang and Guna Lakshminarayanan and Hakan Inan and Hamid Shojanazeri and Han Zou and Hannah Wang and Hanwen Zha and Haroun Habeeb and Harrison Rudolph and Helen Suk and Henry Aspegren and Hunter Goldman and Hongyuan Zhan and Ibrahim Damlaj and Igor Molybog and Igor Tufanov and Ilias Leontiadis and Irina-Elena Veliche and Itai Gat and Jake Weissman and James Geboski and James Kohli and Janice Lam and Japhet Asher and Jean-Baptiste Gaya and Jeff Marcus and Jeff Tang and Jennifer Chan and Jenny Zhen and Jeremy Reizenstein and Jeremy Teboul and Jessica Zhong and Jian Jin and Jingyi Yang and Joe Cummings and Jon Carvill and Jon Shepard and Jonathan McPhie and Jonathan Torres and Josh Ginsburg and Junjie Wang and Kai Wu and Kam Hou U and Karan Saxena and Kartikay Khandelwal and Katayoun Zand and Kathy Matosich and Kaushik Veeraraghavan and Kelly Michelena and Keqian Li and Kiran Jagadeesh and Kun Huang and Kunal Chawla and Kyle Huang and Lailin Chen and Lakshya Garg and Lavender A and Leandro Silva and Lee Bell and Lei Zhang and Liangpeng Guo and Licheng Yu and Liron Moshkovich and Luca Wehrstedt and Madian Khabsa and Manav Avalani and Manish Bhatt and Martynas Mankus and Matan Hasson and Matthew Lennie and Matthias Reso and Maxim Groshev and Maxim Naumov and Maya Lathi and Meghan Keneally and Miao Liu and Michael L. Seltzer and Michal Valko and Michelle Restrepo and Mihir Patel and Mik Vyatskov and Mikayel Samvelyan and Mike Clark and Mike Macey and Mike Wang and Miquel Jubert Hermoso and Mo Metanat and Mohammad Rastegari and Munish Bansal and Nandhini Santhanam and Natascha Parks and Natasha White and Navyata Bawa and Nayan Singhal and Nick Egebo and Nicolas Usunier and Nikhil Mehta and Nikolay Pavlovich Laptev and Ning Dong and Norman Cheng and Oleg Chernoguz and Olivia Hart and Omkar Salpekar and Ozlem Kalinli and Parkin Kent and Parth Parekh and Paul Saab and Pavan Balaji and Pedro Rittner and Philip Bontrager and Pierre Roux and Piotr Dollar and Polina Zvyagina and Prashant Ratanchandani and Pritish Yuvraj and Qian Liang and Rachad Alao and Rachel Rodriguez and Rafi Ayub and Raghotham Murthy and Raghu Nayani and Rahul Mitra and Rangaprabhu Parthasarathy and Raymond Li and Rebekkah Hogan and Robin Battey and Rocky Wang and Russ Howes and Ruty Rinott and Sachin Mehta and Sachin Siby and Sai Jayesh Bondu and Samyak Datta and Sara Chugh and Sara Hunt and Sargun Dhillon and Sasha Sidorov and Satadru Pan and Saurabh Mahajan and Saurabh Verma and Seiji Yamamoto and Sharadh Ramaswamy and Shaun Lindsay and Shaun Lindsay and Sheng Feng and Shenghao Lin and Shengxin Cindy Zha and Shishir Patil and Shiva Shankar and Shuqiang Zhang and Shuqiang Zhang and Sinong Wang and Sneha Agarwal and Soji Sajuyigbe and Soumith Chintala and Stephanie Max and Stephen Chen and Steve Kehoe and Steve Satterfield and Sudarshan Govindaprasad and Sumit Gupta and Summer Deng and Sungmin Cho and Sunny Virk and Suraj Subramanian and Sy Choudhury and Sydney Goldman and Tal Remez and Tamar Glaser and Tamara Best and Thilo Koehler and Thomas Robinson and Tianhe Li and Tianjun Zhang and Tim Matthews and Timothy Chou and Tzook Shaked and Varun Vontimitta and Victoria Ajayi and Victoria Montanez and Vijai Mohan and Vinay Satish Kumar and Vishal Mangla and Vlad Ionescu and Vlad Poenaru and Vlad Tiberiu Mihailescu and Vladimir Ivanov and Wei Li and Wenchen Wang and Wenwen Jiang and Wes Bouaziz and Will Constable and Xiaocheng Tang and Xiaojian Wu and Xiaolan Wang and Xilun Wu and Xinbo Gao and Yaniv Kleinman and Yanjun Chen and Ye Hu and Ye Jia and Ye Qi and Yenda Li and Yilin Zhang and Ying Zhang and Yossi Adi and Youngjin Nam and Yu and Wang and Yu Zhao and Yuchen Hao and Yundi Qian and Yunlu Li and Yuzi He and Zach Rait and Zachary DeVito and Zef Rosnbrick and Zhaoduo Wen and Zhenyu Yang and Zhiwei Zhao and Zhiyu Ma},
      year={2024},
      eprint={2407.21783},
      archivePrefix={arXiv},
      primaryClass={cs.AI},
      url={https://arxiv.org/abs/2407.21783}, 
}

@article{guo2024largelanguagemodelbased,
      title={Large Language Model based Multi-Agents: A Survey of Progress and Challenges}, 
      author={Taicheng Guo and Xiuying Chen and Yaqi Wang and Ruidi Chang and Shichao Pei and Nitesh V. Chawla and Olaf Wiest and Xiangliang Zhang},
      year={2024},
      eprint={2402.01680},
      archivePrefix={arXiv},
      primaryClass={cs.CL},
      url={https://arxiv.org/abs/2402.01680}, 
}

@article{tran2025multiagentcollaborationmechanismssurvey,
      title={Multi-Agent Collaboration Mechanisms: A Survey of LLMs}, 
      author={Khanh-Tung Tran and Dung Dao and Minh-Duong Nguyen and Quoc-Viet Pham and Barry O'Sullivan and Hoang D. Nguyen},
      year={2025},
      eprint={2501.06322},
      archivePrefix={arXiv},
      primaryClass={cs.AI},
      url={https://arxiv.org/abs/2501.06322}, 
}

@article{li2024survey_on_llm_based_multi_agent_systems,
	author = {Li, Xinyi and Wang, Sai and Zeng, Siqi and Wu, Yu and Yang, Yi},
	date = {2024/10/08},
	doi = {10.1007/s44336-024-00009-2},
	id = {Li2024},
	isbn = {3005-060X},
	journal = {Vicinagearth},
	number = {1},
	pages = {9},
	title = {A survey on LLM-based multi-agent systems: workflow, infrastructure, and challenges},
	url = {https://doi.org/10.1007/s44336-024-00009-2},
	volume = {1},
	year = {2024}}

@article{kaplan2020scaling,
      title={Scaling Laws for Neural Language Models}, 
      author={Jared Kaplan and Sam McCandlish and Tom Henighan and Tom B. Brown and Benjamin Chess and Rewon Child and Scott Gray and Alec Radford and Jeffrey Wu and Dario Amodei},
      year={2020},
      eprint={2001.08361},
      archivePrefix={arXiv},
      primaryClass={cs.LG},
      url={https://arxiv.org/abs/2001.08361}, 
}

@article{hoffmann2022training,
      title={Training Compute-Optimal Large Language Models}, 
      author={Jordan Hoffmann and Sebastian Borgeaud and Arthur Mensch and Elena Buchatskaya and Trevor Cai and Eliza Rutherford and Diego de Las Casas and Lisa Anne Hendricks and Johannes Welbl and Aidan Clark and Tom Hennigan and Eric Noland and Katie Millican and George van den Driessche and Bogdan Damoc and Aurelia Guy and Simon Osindero and Karen Simonyan and Erich Elsen and Jack W. Rae and Oriol Vinyals and Laurent Sifre},
      year={2022},
      eprint={2203.15556},
      archivePrefix={arXiv},
      primaryClass={cs.CL},
      url={https://arxiv.org/abs/2203.15556}, 
}

@article{Bothmann:2020ywa,
    author = "Bothmann, Enrico and Jan{\ss}en, Timo and Knobbe, Max and Schmale, Tobias and Schumann, Steffen",
    title = "{Exploring phase space with Neural Importance Sampling}",
    eprint = "2001.05478",
    archivePrefix = "arXiv",
    primaryClass = "hep-ph",
    reportNumber = "MCNET-20-02, MCNET-20-01",
    doi = "10.21468/SciPostPhys.8.4.069",
    journal = "SciPost Phys.",
    volume = "8",
    number = "4",
    pages = "069",
    year = "2020"
}

@article{Gao:2020zvv,
    author = {Gao, Christina and H{\"o}che, Stefan and Isaacson, Joshua and Krause, Claudius and Schulz, Holger},
    title = "{Event Generation with Normalizing Flows}",
    eprint = "2001.10028",
    archivePrefix = "arXiv",
    primaryClass = "hep-ph",
    reportNumber = "FERMILAB-PUB-20-009-SCD-T, MCNET-20-03",
    doi = "10.1103/PhysRevD.101.076002",
    journal = "Phys. Rev. D",
    volume = "101",
    number = "7",
    pages = "076002",
    year = "2020"
}

@article{Ilten:2022jfm,
    author = "Ilten, Phil and Menzo, Tony and Youssef, Ahmed and Zupan, Jure",
    title = "{Modeling hadronization using machine learning}",
    eprint = "2203.04983",
    archivePrefix = "arXiv",
    primaryClass = "hep-ph",
    doi = "10.21468/SciPostPhys.14.3.027",
    journal = "SciPost Phys.",
    volume = "14",
    number = "3",
    pages = "027",
    year = "2023"
}

@article{Bierlich:2023zzd,
    author = "Bierlich, Christian and Ilten, Phil and Menzo, Tony and Mrenna, Stephen and Szewc, Manuel and Wilkinson, Michael K. and Youssef, Ahmed and Zupan, Jure",
    title = "{Towards a data-driven model of hadronization using normalizing flows}",
    eprint = "2311.09296",
    archivePrefix = "arXiv",
    primaryClass = "hep-ph",
    reportNumber = "FERMILAB-PUB-23-698-CSAID",
    doi = "10.21468/SciPostPhys.17.2.045",
    journal = "SciPost Phys.",
    volume = "17",
    number = "2",
    pages = "045",
    year = "2024"
}

@article{Bierlich:2024xzg,
    author = "Bierlich, Christian and Ilten, Phil and Menzo, Tony and Mrenna, Stephen and Szewc, Manuel and Wilkinson, Michael K. and Youssef, Ahmed and Zupan, Jure",
    title = "{Describing hadronization via histories and observables for Monte-Carlo event reweighting}",
    eprint = "2410.06342",
    archivePrefix = "arXiv",
    primaryClass = "hep-ph",
    reportNumber = "FERMILAB-PUB-23-414-CSAID",
    doi = "10.21468/SciPostPhys.18.2.054",
    journal = "SciPost Phys.",
    volume = "18",
    number = "2",
    pages = "054",
    year = "2025"
}

@article{Assi:2025avy,
    author = "Assi, Beno{\^\i}t and Bierlich, Christian and Ilten, Phil and Menzo, Tony and Mrenna, Stephen and Szewc, Manuel and Wilkinson, Michael K. and Youssef, Ahmed and Zupan, Jure",
    title = "{Characterizing the hadronization of parton showers using the HOMER method}",
    eprint = "2503.05667",
    archivePrefix = "arXiv",
    primaryClass = "hep-ph",
    reportNumber = "FERMILAB-PUB-25-0133-CSAID",
    doi = "10.21468/SciPostPhys.19.5.125",
    journal = "SciPost Phys.",
    volume = "19",
    pages = "125",
    year = "2025"
}

@article{Butter:2025wxn,
    author = "Butter, Anja and others",
    title = "{Iterative HOMER with uncertainties}",
    eprint = "2509.03592",
    archivePrefix = "arXiv",
    primaryClass = "hep-ph",
    reportNumber = "FERMILAB-PUB-25-0579-CSAID",
    month = "9",
    year = "2025"
}

@article{NNPDF:2021njg,
    author = "Ball, Richard D. and others",
    collaboration = "NNPDF",
    title = "{The path to proton structure at 1{\%} accuracy}",
    eprint = "2109.02653",
    archivePrefix = "arXiv",
    primaryClass = "hep-ph",
    reportNumber = "Edinburgh 2021/12, Nikhef-2021-013, TIF-UNIMI-2021-11",
    doi = "10.1140/epjc/s10052-022-10328-7",
    journal = "Eur. Phys. J. C",
    volume = "82",
    number = "5",
    pages = "428",
    year = "2022"
}

@article{NNPDF:2021uiq,
    author = "Ball, Richard D. and others",
    collaboration = "NNPDF",
    title = "{An open-source machine learning framework for global analyses of parton distributions}",
    eprint = "2109.02671",
    archivePrefix = "arXiv",
    primaryClass = "hep-ph",
    reportNumber = "Edinburgh 2021/13, Nikhef-2021-020, TIF-UNIMI-2021-12",
    doi = "10.1140/epjc/s10052-021-09747-9",
    journal = "Eur. Phys. J. C",
    volume = "81",
    number = "10",
    pages = "958",
    year = "2021"
}

@article{Butter:2023fov,
    author = "Butter, Anja and Huetsch, Nathan and Palacios Schweitzer, Sofia and Plehn, Tilman and Sorrenson, Peter and Spinner, Jonas",
    title = "{Jet diffusion versus JetGPT {\textendash} Modern networks for the LHC}",
    eprint = "2305.10475",
    archivePrefix = "arXiv",
    primaryClass = "hep-ph",
    doi = "10.21468/SciPostPhysCore.8.1.026",
    journal = "SciPost Phys. Core",
    volume = "8",
    pages = "026",
    year = "2025"
}

@article{Antcheva:2009zz,
    author = "Antcheva, I. and others",
    title = "{ROOT: A C++ framework for petabyte data storage, statistical analysis and visualization}",
    eprint = "1508.07749",
    archivePrefix = "arXiv",
    primaryClass = "physics.data-an",
    reportNumber = "FERMILAB-PUB-09-661-CD",
    doi = "10.1016/j.cpc.2009.08.005",
    journal = "Comput. Phys. Commun.",
    volume = "180",
    pages = "2499--2512",
    year = "2009"
}

@article{Alwall:2006yp,
    author = "Alwall, J. and others",
    title = "{A Standard format for Les Houches event files}",
    eprint = "hep-ph/0609017",
    archivePrefix = "arXiv",
    reportNumber = "FERMILAB-PUB-06-337-T, CERN-LCGAPP-2006-03",
    doi = "10.1016/j.cpc.2006.11.010",
    journal = "Comput. Phys. Commun.",
    volume = "176",
    pages = "300--304",
    year = "2007"
}

@article{Bakshi:2025fgx,
    author = "Bakshi, S. D. and others",
    title = "{ArgoLOOM: agentic AI for fundamental physics from quarks to cosmos}",
    eprint = "2510.02426",
    archivePrefix = "arXiv",
    primaryClass = "hep-ph",
    reportNumber = "ANL-199516",
    month = "10",
    year = "2025"
}

@article{chemcrow,
    author = "M. Bran, Andres and Cox, Sam and Schilter, Oliver and Baldassari, Carlo and White, Andrew D. and Schwaller, Philippe",
    title = "{ChemCrow: Augmenting large-language models with chemistry tools}",
    eprint = "2304.05376",
    archivePrefix = "arXiv",
    primaryClass = "physics.chem-ph",
    month = "4",
    year = "2023"
}

@article{coscientist,
    author = "Boiko, Daniil A. and MacKnight, Robert and Kline, Ben and Gomes, Gabe",
    title = "{Autonomous chemical research with large language models}",
    journal = "Nature",
    volume = "624",
    number = "7992",
    pages = "570--578",
    year = "2023"
}

@article{metagpt,
    author = "Hong, Sirui and others",
    title = "{MetaGPT: Meta Programming for a Multi-Agent Collaborative Framework}",
    eprint = "2308.00352",
    archivePrefix = "arXiv",
    primaryClass = "cs.AI",
    month = "8",
    year = "2023"
}

@article{sweagent,
    author = "Yang, John and Jimenez, Carlos E. and Wettig, Alexander and Lieret, Kilian and Yao, Shunyu and Narasimhan, Karthik and Press, Ofir",
    title = "{SWE-agent: Agent-Computer Interfaces Enable Automated Software Engineering}",
    eprint = "2405.15793",
    archivePrefix = "arXiv",
    primaryClass = "cs.SE",
    month = "5",
    year = "2024"
}

@article{econagent,
    author = "Li, Nian and Gao, Chen and Li, Mingyu and Li, Yong and Liao, Qingmin",
    title = "{EconAgent: Large Language Model-Empowered Agents for Simulating Macroeconomic Activities}",
    eprint = "2310.10436",
    archivePrefix = "arXiv",
    primaryClass = "cs.AI",
    month = "10",
    year = "2023"
}

@article{Laverick:2024mas,
    author = "Laverick, Andrew and Surrao, Kristen and Zubeldia, Inigo and Bolliet, Boris and Cranmer, Miles and Lewis, Antony and Sherwin, Blake and Lesgourgues, Julien",
    title = "{Multi-Agent System for Cosmological Parameter Analysis}",
    eprint = "2412.00431",
    archivePrefix = "arXiv",
    primaryClass = "astro-ph.IM",
    month = "12",
    year = "2024"
}

@article{Moss:2025aicosmologist,
    author = "Moss, Adam",
    title = "{The AI Cosmologist I: An Agentic System for Automated Data Analysis}",
    eprint = "2504.03424",
    archivePrefix = "arXiv",
    primaryClass = "astro-ph.IM",
    month = "4",
    year = "2025"
}

@article{Xu:2025cmbagent,
    author = "Xu, Licong and Sarkar, Milind and Lonappan, Anto I. and others",
    title = "{Open Source Planning \& Control System with Language Agents for Autonomous Scientific Discovery}",
    eprint = "2507.07257",
    archivePrefix = "arXiv",
    primaryClass = "astro-ph.IM",
    month = "7",
    year = "2025"
}

@article{Diefenbacher:2025aod,
    author = "Diefenbacher, Sascha and Hallin, Anna and Kasieczka, Gregor and Kramer, Michael and Lauscher, Anne and others",
    title = "{Agents of Discovery}",
    eprint = "2509.08535",
    archivePrefix = "arXiv",
    primaryClass = "hep-ph",
    month = "9",
    year = "2025"
}

@article{GendreauDistler:2025hep,
    author = "Gendreau-Distler, Eli and Ho, Joshua and Kim, Dongwon and Le Pottier, Luc Tomas and Wang, Haichen and Yang, Chengxi",
    title = "{Automating High Energy Physics Data Analysis with LLM-Powered Agents}",
    eprint = "2512.07785",
    archivePrefix = "arXiv",
    primaryClass = "hep-ex",
    month = "12",
    year = "2025"
}

@article{Qu:2019gqs,
    author = "Qu, Huilin and Gouskos, Loukas",
    title = "{ParticleNet: Jet Tagging via Particle Clouds}",
    eprint = "1902.08570",
    archivePrefix = "arXiv",
    primaryClass = "hep-ph",
    doi = "10.1103/PhysRevD.101.056019",
    journal = "Phys. Rev. D",
    volume = "101",
    number = "5",
    pages = "056019",
    year = "2020"
}

@article{Qu:2022mxj,
    author = "Qu, Huilin and Li, Congqiao and Qian, Sitian",
    title = "{Particle Transformer for Jet Tagging}",
    eprint = "2202.03772",
    archivePrefix = "arXiv",
    primaryClass = "hep-ph",
    month = "2",
    year = "2022"
}

@article{Butter:2022rso,
    author = "Badger, Simon and others",
    editor = "Butter, Anja and Plehn, Tilman and Schumann, Steffen",
    title = "{Machine learning and LHC event generation}",
    eprint = "2203.07460",
    archivePrefix = "arXiv",
    primaryClass = "hep-ph",
    reportNumber = "FERMILAB-PUB-22-126-T",
    doi = "10.21468/SciPostPhys.14.4.079",
    journal = "SciPost Phys.",
    volume = "14",
    number = "4",
    pages = "079",
    year = "2023"
}

@article{Birk:2024knn,
    author = "Birk, Joschka and Hallin, Anna and Kasieczka, Gregor",
    title = "{OmniJet-{\ensuremath{\alpha}}: the first cross-task foundation model for particle physics}",
    eprint = "2403.05618",
    archivePrefix = "arXiv",
    primaryClass = "hep-ph",
    doi = "10.1088/2632-2153/ad66ad",
    journal = "Mach. Learn. Sci. Tech.",
    volume = "5",
    number = "3",
    pages = "035031",
    year = "2024"
}

@article{Golling:2024abg,
    author = "Golling, Tobias and Heinrich, Lukas and Kagan, Michael and Klein, Samuel and Leigh, Matthew and Osadchy, Margarita and Raine, John Andrew",
    title = "{Masked particle modeling on sets: towards self-supervised high energy physics foundation models}",
    eprint = "2401.13537",
    archivePrefix = "arXiv",
    primaryClass = "hep-ph",
    doi = "10.1088/2632-2153/ad64a8",
    journal = "Mach. Learn. Sci. Tech.",
    volume = "5",
    number = "3",
    pages = "035074",
    year = "2024"
}

@misc{Bardhan:2025icr,
    author = "Bardhan, Jai and Agrawal, Radhikesh and Tilak, Abhiram and Neeraj, Cyrin and Mitra, Subhadip",
    title = "{HEP-JEPA: A foundation model for collider physics using joint embedding predictive architecture}",
    eprint = "2502.03933",
    archivePrefix = "arXiv",
    primaryClass = "cs.LG",
    month = "2",
    year = "2025"
}

@inproceedings{ExaTrkX:2020nyf,
    author = "Ju, Xiangyang and others",
    collaboration = "Exa.TrkX",
    title = "{Graph Neural Networks for Particle Reconstruction in High Energy Physics detectors}",
    booktitle = "{33rd Annual Conference on Neural Information Processing Systems}",
    eprint = "2003.11603",
    archivePrefix = "arXiv",
    primaryClass = "physics.ins-det",
    reportNumber = "FERMILAB-CONF-20-163-PPD-QIS-SCD",
    month = "3",
    year = "2020"
}

@article{Iiyama:2020wap,
    author = "Iiyama, Yutaro and others",
    title = "{Distance-Weighted Graph Neural Networks on FPGAs for Real-Time Particle Reconstruction in High Energy Physics}",
    eprint = "2008.03601",
    archivePrefix = "arXiv",
    primaryClass = "physics.ins-det",
    reportNumber = "FERMILAB-PUB-20-405-E-SCD",
    doi = "10.3389/fdata.2020.598927",
    journal = "Front. Big Data",
    volume = "3",
    pages = "598927",
    year = "2020"
}

@article{Duarte:2019fta,
    author = "Duarte, Javier and others",
    title = "{FPGA-accelerated machine learning inference as a service for particle physics computing}",
    eprint = "1904.08986",
    archivePrefix = "arXiv",
    primaryClass = "physics.data-an",
    reportNumber = "FERMILAB-PUB-19-170-CD-CMS-E-ND",
    doi = "10.1007/s41781-019-0027-2",
    journal = "Comput. Softw. Big Sci.",
    volume = "3",
    number = "1",
    pages = "13",
    year = "2019"
}

@article{Aarrestad:2021oeb,
    author = "Aarrestad, Thea and others",
    title = "{The Dark Machines Anomaly Score Challenge: Benchmark Data and Model Independent Event Classification for the Large Hadron Collider}",
    eprint = "2105.14027",
    archivePrefix = "arXiv",
    primaryClass = "hep-ph",
    reportNumber = "FERMILAB-PUB-21-285-CMS",
    doi = "10.21468/SciPostPhys.12.1.043",
    journal = "SciPost Phys.",
    volume = "12",
    number = "1",
    pages = "043",
    year = "2022"
}

@article{Kasieczka:2021xcg,
    author = "Kasieczka, Gregor and others",
    title = "{The LHC Olympics 2020 a community challenge for anomaly detection in high energy physics}",
    eprint = "2101.08320",
    archivePrefix = "arXiv",
    primaryClass = "hep-ph",
    doi = "10.1088/1361-6633/ac36b9",
    journal = "Rept. Prog. Phys.",
    volume = "84",
    number = "12",
    pages = "124201",
    year = "2021"
}

@article{Andreassen:2019cjw,
    author = "Andreassen, Anders and Komiske, Patrick T. and Metodiev, Eric M. and Nachman, Benjamin and Thaler, Jesse",
    title = "{OmniFold: A Method to Simultaneously Unfold All Observables}",
    eprint = "1911.09107",
    archivePrefix = "arXiv",
    primaryClass = "hep-ph",
    reportNumber = "MIT-CTP 5155",
    doi = "10.1103/PhysRevLett.124.182001",
    journal = "Phys. Rev. Lett.",
    volume = "124",
    number = "18",
    pages = "182001",
    year = "2020"
}

@article{Bellagente:2020piv,
    author = {Bellagente, Marco and Butter, Anja and Kasieczka, Gregor and Plehn, Tilman and Rousselot, Armand and Winterhalder, Ramon and Ardizzone, Lynton and K{\"o}the, Ullrich},
    title = "{Invertible Networks or Partons to Detector and Back Again}",
    eprint = "2006.06685",
    archivePrefix = "arXiv",
    primaryClass = "hep-ph",
    doi = "10.21468/SciPostPhys.9.5.074",
    journal = "SciPost Phys.",
    volume = "9",
    pages = "074",
    year = "2020"
}

@article{Albertsson:2018maf,
    author = "Albertsson, Kim and others",
    title = "{Machine Learning in High Energy Physics Community White Paper}",
    eprint = "1807.02876",
    archivePrefix = "arXiv",
    primaryClass = "physics.comp-ph",
    reportNumber = "FERMILAB-PUB-18-318-CD-DI-PPD",
    doi = "10.1088/1742-6596/1085/2/022008",
    journal = "J. Phys. Conf. Ser.",
    volume = "1085",
    number = "2",
    pages = "022008",
    year = "2018"
}

@article{Guest:2018yhq,
    author = "Guest, Dan and Cranmer, Kyle and Whiteson, Daniel",
    title = "{Deep Learning and its Application to LHC Physics}",
    eprint = "1806.11484",
    archivePrefix = "arXiv",
    primaryClass = "hep-ex",
    doi = "10.1146/annurev-nucl-101917-021019",
    journal = "Ann. Rev. Nucl. Part. Sci.",
    volume = "68",
    pages = "161--181",
    year = "2018"
}

@article{Zhang:2024kws,
    author = "Zhang, Zhengde and others",
    title = "{Xiwu: A Basis Flexible and Learnable LLM for High Energy Physics}",
    eprint = "2404.08001",
    archivePrefix = "arXiv",
    primaryClass = "hep-ph",
    month = "4",
    year = "2024"
}

@article{Atif:2025mkh,
    author = "Atif, Mohammad and Chopra, Kriti and Kilic, Ozgur and Wang, Tianle and Dong, Zhihua and Leggett, Charles and Lin, Meifeng and Calafiura, Paolo and Habib, Salman",
    title = "{CelloAI: Leveraging Large Language Models for HPC Software Development in High Energy Physics}",
    eprint = "2508.16713",
    archivePrefix = "arXiv",
    primaryClass = "cs.SE",
    month = "8",
    year = "2025"
}

@article{Dorsner:2016wpm,
    author = "Dor{\v{s}}ner, I. and Fajfer, S. and Greljo, A. and Kamenik, J. F. and Ko{\v{s}}nik, N.",
    title = "{Physics of leptoquarks in precision experiments and at particle colliders}",
    eprint = "1603.04993",
    archivePrefix = "arXiv",
    primaryClass = "hep-ph",
    doi = "10.1016/j.physrep.2016.06.001",
    journal = "Phys. Rept.",
    volume = "641",
    pages = "1--68",
    year = "2016"
}

@article{openai2025gptoss120bgptoss20bmodel,
      title={gpt-oss-120b and gpt-oss-20b Model Card}, 
      author={OpenAI and : and Sandhini Agarwal and Lama Ahmad and Jason Ai and Sam Altman and Andy Applebaum and Edwin Arbus and Rahul K. Arora and Yu Bai and Bowen Baker and Haiming Bao and Boaz Barak and Ally Bennett and Tyler Bertao and Nivedita Brett and Eugene Brevdo and Greg Brockman and Sebastien Bubeck and Che Chang and Kai Chen and Mark Chen and Enoch Cheung and Aidan Clark and Dan Cook and Marat Dukhan and Casey Dvorak and Kevin Fives and Vlad Fomenko and Timur Garipov and Kristian Georgiev and Mia Glaese and Tarun Gogineni and Adam Goucher and Lukas Gross and Katia Gil Guzman and John Hallman and Jackie Hehir and Johannes Heidecke and Alec Helyar and Haitang Hu and Romain Huet and Jacob Huh and Saachi Jain and Zach Johnson and Chris Koch and Irina Kofman and Dominik Kundel and Jason Kwon and Volodymyr Kyrylov and Elaine Ya Le and Guillaume Leclerc and James Park Lennon and Scott Lessans and Mario Lezcano-Casado and Yuanzhi Li and Zhuohan Li and Ji Lin and Jordan Liss and Lily and Liu and Jiancheng Liu and Kevin Lu and Chris Lu and Zoran Martinovic and Lindsay McCallum and Josh McGrath and Scott McKinney and Aidan McLaughlin and Song Mei and Steve Mostovoy and Tong Mu and Gideon Myles and Alexander Neitz and Alex Nichol and Jakub Pachocki and Alex Paino and Dana Palmie and Ashley Pantuliano and Giambattista Parascandolo and Jongsoo Park and Leher Pathak and Carolina Paz and Ludovic Peran and Dmitry Pimenov and Michelle Pokrass and Elizabeth Proehl and Huida Qiu and Gaby Raila and Filippo Raso and Hongyu Ren and Kimmy Richardson and David Robinson and Bob Rotsted and Hadi Salman and Suvansh Sanjeev and Max Schwarzer and D. Sculley and Harshit Sikchi and Kendal Simon and Karan Singhal and Yang Song and Dane Stuckey and Zhiqing Sun and Philippe Tillet and Sam Toizer and Foivos Tsimpourlas and Nikhil Vyas and Eric Wallace and Xin Wang and Miles Wang and Olivia Watkins and Kevin Weil and Amy Wendling and Kevin Whinnery and Cedric Whitney and Hannah Wong and Lin Yang and Yu Yang and Michihiro Yasunaga and Kristen Ying and Wojciech Zaremba and Wenting Zhan and Cyril Zhang and Brian Zhang and Eddie Zhang and Shengjia Zhao},
      year={2025},
      eprint={2508.10925},
      archivePrefix={arXiv},
      primaryClass={cs.CL},
      url={https://arxiv.org/abs/2508.10925}, 
}

@article{Ask:2012sm,
    author = "Ask, Stefan and others",
    title = "{From Lagrangians to Events: Computer Tutorial at the MC4BSM-2012 Workshop}",
    eprint = "1209.0297",
    archivePrefix = "arXiv",
    primaryClass = "hep-ph",
    reportNumber = "CERN-PH-TH-2012-280",
    month = "9",
    year = "2012"
}

@article{Alwall:2014hca,
    author = "Alwall, J. and Frederix, R. and Frixione, S. and Hirschi, V. and Maltoni, F. and Mattelaer, O. and Shao, H. -S. and Stelzer, T. and Torrielli, P. and Zaro, M.",
    title = "{The automated computation of tree-level and next-to-leading order differential cross sections, and their matching to parton shower simulations}",
    eprint = "1405.0301",
    archivePrefix = "arXiv",
    primaryClass = "hep-ph",
    reportNumber = "CERN-PH-TH-2014-064, CP3-14-18, LPN14-066, MCNET-14-09, ZU-TH-14-14",
    doi = "10.1007/JHEP07(2014)079",
    journal = "JHEP",
    volume = "07",
    pages = "079",
    year = "2014"
}

@article{Cacciari:2011ma,
    author = "Cacciari, Matteo and Salam, Gavin P. and Soyez, Gregory",
    title = "{FastJet User Manual}",
    eprint = "1111.6097",
    archivePrefix = "arXiv",
    primaryClass = "hep-ph",
    reportNumber = "CERN-PH-TH-2011-297",
    doi = "10.1140/epjc/s10052-012-1896-2",
    journal = "Eur. Phys. J. C",
    volume = "72",
    pages = "1896",
    year = "2012"
}

@article{Bierlich:2022pfr,
    author = "Bierlich, Christian and others",
    title = "{A comprehensive guide to the physics and usage of PYTHIA 8.3}",
    eprint = "2203.11601",
    archivePrefix = "arXiv",
    primaryClass = "hep-ph",
    reportNumber = "LU-TP 22-16, MCNET-22-04, FERMILAB-PUB-22-227-SCD",
    doi = "10.21468/SciPostPhysCodeb.8",
    journal = "SciPost Phys. Codeb.",
    volume = "2022",
    pages = "8",
    year = "2022"
}

@article{Alloul:2013bka,
    author = "Alloul, Adam and Christensen, Neil D. and Degrande, C{\'e}line and Duhr, Claude and Fuks, Benjamin",
    title = "{FeynRules  2.0 - A complete toolbox for tree-level phenomenology}",
    eprint = "1310.1921",
    archivePrefix = "arXiv",
    primaryClass = "hep-ph",
    reportNumber = "CERN-PH-TH-2013-239, MCNET-13-14, IPPP-13-71, DCPT-13-142, PITT-PACC-1308",
    doi = "10.1016/j.cpc.2014.04.012",
    journal = "Comput. Phys. Commun.",
    volume = "185",
    pages = "2250--2300",
    year = "2014"
}

@article{Alwall:2011uj,
    author = "Alwall, Johan and Herquet, Michel and Maltoni, Fabio and Mattelaer, Olivier and Stelzer, Tim",
    title = "{MadGraph 5 : Going Beyond}",
    eprint = "1106.0522",
    archivePrefix = "arXiv",
    primaryClass = "hep-ph",
    reportNumber = "FERMILAB-PUB-11-448-T",
    doi = "10.1007/JHEP06(2011)128",
    journal = "JHEP",
    volume = "06",
    pages = "128",
    year = "2011"
}

@article{Bewick:2023tfi,
    author = "Bewick, Gavin and others",
    title = "{Herwig 7.3 release note}",
    eprint = "2312.05175",
    archivePrefix = "arXiv",
    primaryClass = "hep-ph",
    reportNumber = "CERN-TH-2023-223, HERWIG-2023-01, KA-TP-28-2023, MCnet-23-19, IPPP/23/66",
    doi = "10.1140/epjc/s10052-024-13211-9",
    journal = "Eur. Phys. J. C",
    volume = "84",
    number = "10",
    pages = "1053",
    year = "2024"
}

@article{deFavereau:2013fsa,
    author = "de Favereau, J. and Delaere, C. and Demin, P. and Giammanco, A. and Lema{\^\i}tre, V. and Mertens, A. and Selvaggi, M.",
    collaboration = "DELPHES 3",
    title = "{DELPHES 3, A modular framework for fast simulation of a generic collider experiment}",
    eprint = "1307.6346",
    archivePrefix = "arXiv",
    primaryClass = "hep-ex",
    doi = "10.1007/JHEP02(2014)057",
    journal = "JHEP",
    volume = "02",
    pages = "057",
    year = "2014"
}

@article{Allison:2016lfl,
    author = "Allison, J. and others",
    title = "{Recent developments in Geant4}",
    reportNumber = "FERMILAB-PUB-16-447-CD",
    doi = "10.1016/j.nima.2016.06.125",
    journal = "Nucl. Instrum. Meth. A",
    volume = "835",
    pages = "186--225",
    year = "2016"
}

@article{Bierlich:2019rhm,
    author = "Bierlich, Christian and others",
    title = "{Robust Independent Validation of Experiment and Theory: Rivet version 3}",
    eprint = "1912.05451",
    archivePrefix = "arXiv",
    primaryClass = "hep-ph",
    reportNumber = "MCnet-19-26",
    doi = "10.21468/SciPostPhys.8.2.026",
    journal = "SciPost Phys.",
    volume = "8",
    pages = "026",
    year = "2020"
}

@article{Isaacson:2022cwh,
    author = "Isaacson, Joshua and Jay, William I. and Lovato, Alessandro and Machado, Pedro A. N. and Rocco, Noemi",
    title = "{Introducing a novel event generator for electron-nucleus and neutrino-nucleus scattering}",
    eprint = "2205.06378",
    archivePrefix = "arXiv",
    primaryClass = "hep-ph",
    reportNumber = "FERMILAB-PUB-22-411-T, MIT-CTP/5428",
    doi = "10.1103/PhysRevD.107.033007",
    journal = "Phys. Rev. D",
    volume = "107",
    number = "3",
    pages = "033007",
    year = "2023"
}

@article{Andreopoulos:2009rq,
    author = "Andreopoulos, C. and others",
    title = "{The GENIE Neutrino Monte Carlo Generator}",
    eprint = "0905.2517",
    archivePrefix = "arXiv",
    primaryClass = "hep-ph",
    reportNumber = "FERMILAB-PUB-09-418-CD",
    doi = "10.1016/j.nima.2009.12.009",
    journal = "Nucl. Instrum. Meth. A",
    volume = "614",
    pages = "87--104",
    year = "2010"
}

@article{jiang2025agenticscimlcollaborativemultiagentsystems,
      title={AgenticSciML: Collaborative Multi-Agent Systems for Emergent Discovery in Scientific Machine Learning}, 
      author={Qile Jiang and George Karniadakis},
      year={2025},
      eprint={2511.07262},
      archivePrefix={arXiv},
      primaryClass={cs.AI},
      url={https://arxiv.org/abs/2511.07262}, 
}

@article{wei2025aiscienceagenticscience,
      title={From AI for Science to Agentic Science: A Survey on Autonomous Scientific Discovery}, 
      author={Jiaqi Wei and Yuejin Yang and Xiang Zhang and Yuhan Chen and Xiang Zhuang and Zhangyang Gao and Dongzhan Zhou and Guangshuai Wang and Zhiqiang Gao and Juntai Cao and Zijie Qiu and Ming Hu and Chenglong Ma and Shixiang Tang and Junjun He and Chunfeng Song and Xuming He and Qiang Zhang and Chenyu You and Shuangjia Zheng and Ning Ding and Wanli Ouyang and Nanqing Dong and Yu Cheng and Siqi Sun and Lei Bai and Bowen Zhou},
      year={2025},
      eprint={2508.14111},
      archivePrefix={arXiv},
      primaryClass={cs.LG},
      url={https://arxiv.org/abs/2508.14111}, 
}

@misc{MC4BSM,
  title = {Monte Carlo Tools for Beyond the Standard Model Physics},
  howpublished = {\url{https://theory.fnal.gov/mc4bsm/}},
  note = {Accessed: 2025-12-10}
}

@misc{DiscoveryScience,
  title = {Genesis Mission: A National Mission to Accelerate Science Through Artificial Intelligence},
  howpublished = {\url{https://genesis.energy.gov/}},
  note = {Accessed: 2025-12-11}
}

@article{Buckley:2019xhk,
    author = {Buckley, Andy and Ilten, Philip and Konstantinov, Dmitri and L{\"o}nnblad, Leif and Monk, James and Pokorski, Witold and Przedzinski, Tomasz and Verbytskyi, Andrii},
    title = "{The HepMC3 event record library for Monte Carlo event generators}",
    eprint = "1912.08005",
    archivePrefix = "arXiv",
    primaryClass = "hep-ph",
    reportNumber = "MPP-2019-258, MCNET-19-27, LU-TP 19-58",
    doi = "10.1016/j.cpc.2020.107310",
    journal = "Comput. Phys. Commun.",
    volume = "260",
    pages = "107310",
    year = "2021"
}

@article{Dong:2022trn,
    author = "Dong, Zhongtian and Kong, Kyoungchul and Matchev, Konstantin T. and Matcheva, Katia",
    title = "{Is the machine smarter than the theorist: Deriving formulas for particle kinematics with symbolic regression}",
    eprint = "2211.08420",
    archivePrefix = "arXiv",
    primaryClass = "hep-ph",
    doi = "10.1103/PhysRevD.107.055018",
    journal = "Phys. Rev. D",
    volume = "107",
    number = "5",
    pages = "055018",
    year = "2023"
}

@article{Alnuqaydan:2022ncd,
    author = "Alnuqaydan, Abdulhakim and Gleyzer, Sergei and Prosper, Harrison",
    title = "{SYMBA: symbolic computation of squared amplitudes in high energy physics with machine learning}",
    eprint = "2206.08901",
    archivePrefix = "arXiv",
    primaryClass = "hep-ph",
    doi = "10.1088/2632-2153/acb2b2",
    journal = "Mach. Learn. Sci. Tech.",
    volume = "4",
    number = "1",
    pages = "015007",
    year = "2023"
}

@article{Matchev:2024ash,
    author = "Matchev, Konstantin T. and Matcheva, Katia and Ramond, Pierre and Verner, Sarunas",
    title = "{Exploring the truth and beauty of theory landscapes with machine learning}",
    eprint = "2401.11513",
    archivePrefix = "arXiv",
    primaryClass = "hep-ph",
    doi = "10.1016/j.physletb.2024.138941",
    journal = "Phys. Lett. B",
    volume = "856",
    pages = "138941",
    year = "2024"
}

@article{Wojcik:2024lfy,
    author = "Wojcik, George N. and Eu, Shu Tian and Everett, Lisa L.",
    title = "{Graph reinforcement learning for exploring model spaces beyond the standard model}",
    eprint = "2407.07203",
    archivePrefix = "arXiv",
    primaryClass = "hep-ph",
    doi = "10.1103/PhysRevD.111.035007",
    journal = "Phys. Rev. D",
    volume = "111",
    number = "3",
    pages = "035007",
    year = "2025"
}

@article{Kawai:2024pws,
    author = "Kawai, Shinsuke and Okada, Nobuchika",
    title = "{Truth, beauty, and goodness in grand unification: A machine learning approach}",
    eprint = "2411.06718",
    archivePrefix = "arXiv",
    primaryClass = "hep-ph",
    doi = "10.1016/j.physletb.2024.139221",
    journal = "Phys. Lett. B",
    volume = "860",
    pages = "139221",
    year = "2025"
}

@article{Koay:2025bmu,
    author = "Koay, Yong Sheng and Enberg, Rikard and Moretti, Stefano and Camargo-Molina, Eliel",
    title = "{Generating particle physics Lagrangians with transformers}",
    eprint = "2501.09729",
    archivePrefix = "arXiv",
    primaryClass = "cs.LG",
    month = "1",
    year = "2025"
}

@article{Sherpa:2024mfk,
    author = "Bothmann, Enrico and others",
    collaboration = "Sherpa",
    title = "{Event generation with Sherpa 3}",
    eprint = "2410.22148",
    archivePrefix = "arXiv",
    primaryClass = "hep-ph",
    reportNumber = "IPPP/24/67, LTH-1385, FERMILAB-PUB-24-0748-T, ZU-TH 51/24, MCNET-24-17, CERN-TH-2024-171",
    doi = "10.1007/JHEP12(2024)156",
    journal = "JHEP",
    volume = "12",
    pages = "156",
    year = "2024"
}

@inproceedings{kulal2019spoc,
  title={{SPoC}: Search-based Pseudocode to Code},
  author={Kulal, Sumith and Pasupat, Panupong and Chandra, Kartik and Lee, Mina and Padon, Oded and Aiken, Alex and Liang, Percy},
  booktitle={Advances in Neural Information Processing Systems (NeurIPS)},
  year={2019}
}

@article{chen2021evaluating,
  title={Evaluating Large Language Models Trained on Code},
  author={Chen, Mark and Tworek, Jerry and Jun, Heewoo and others},
  journal={arXiv preprint arXiv:2107.03374},
  year={2021}
}

@article{liu2024stablereasoning,
  title={Are Your {LLM}s Capable of Stable Reasoning?},
  author={Liu, Junnan and others},
  eprint="2412.13147",
  year={2024}
}

@misc{HEPTAPOD:benchmarking,
  author = {Menzo, Tony and Roman, Alexander and Gleyzer, Sergei and Matchev, Konstantin and Fleming, George T. and Mrenna, Stephen},
  howpublished = {in preparation}
}
